\documentclass[%
 reprint,
superscriptaddress,
longbibliography,
bibnotes,
 amsmath,amssymb,
 aps,
pre,
floatfix,
]{revtex4-2}

\usepackage{graphicx}
\graphicspath{ {./Figures/} }

\usepackage{dcolumn}
\usepackage{bm}
\usepackage[normalem]{ulem}
\usepackage{xcolor}
\definecolor{OliveGreen}{rgb}{0,0.6,0}
\definecolor{Auburn}{rgb}{0.43, 0.21, 0.1}
\definecolor{blackViolet}{rgb}{0.54, 0.17, 0.89}
\definecolor{HokieOrange}{RGB}{232, 119, 34}
\definecolor{HokieMaroon}{RGB}{134, 31, 65}

\definecolor{HokieMaroon}{RGB}{134, 31, 65}

\usepackage{appendix}
\usepackage{chngcntr}
\usepackage{etoolbox}
\usepackage{lipsum}
\usepackage{graphicx}
\usepackage{mwe}
\usepackage{ulem}

\AtBeginEnvironment{appendices}{%
\addcontentsline{toc}{section}{Appendices}
\counterwithin{equation}{section}
\counterwithin{table}{section}
}

\begin{document}


\title{Predicting 
Non-Equilibrium Folding Behavior of Polymer Chains using the Steepest-Entropy-Ascent Quantum Thermodynamic Framework}

\author{Jared McDonald}
\altaffiliation{jmcdonald@vt.edu (J. McDonald)}
\affiliation{Materials Science and Engineering Department, Virginia Tech, Blacksburg, VA 24061, USA}
\author{Michael R. von Spakovsky}
\altaffiliation{vonspako@vt.edu (M.R. von Spakovsky)}
\affiliation{Mechanical Engineering Department, Virginia Tech, Blacksburg, VA 24061, USA}
\author{William T. Reynolds Jr.}
\altaffiliation{reynolds@vt.edu (W. T. Reynolds Jr.)}
\affiliation{Materials Science and Engineering Department, Virginia Tech, Blacksburg, VA 24061, USA}

\date{2023-01-28}

\begin{abstract}
The steepest-entropy-ascent quantum thermodynamic (SEAQT) framework is used to explore the influence of heating and cooling on polymer chain folding kinetics. The framework predicts how a chain moves from an initial non-equilibrium state to stable equilibrium along a unique thermodynamic path. The thermodynamic state is expressed by occupation probabilities corresponding to the levels of a discrete energy landscape. The landscape is generated using the Replica Exchange Wang-Landau method applied to a polymer chain represented by a sequence of hydrophobic and polar monomers with a simple hydrophobic-polar amino acid model. The chain conformation evolves as energy shifts among the levels of the energy landscape according to the principle of steepest entropy ascent. This principle is implemented via the SEAQT equation of motion. The SEAQT framework has the benefit of providing insight into structural properties under non-equilibrium conditions.  Chain conformations during heating and cooling change continuously without sharp transitions in morphology. The changes are more drastic along non-equilibrium paths than along quasi-equilibrium paths. The SEAQT-predicted kinetics are fitted to rates associated with the experimental intensity profiles of cytochrome c protein folding with Rouse dynamics.
\end{abstract}

                            
\maketitle

\section{Introduction}
There has been considerable interest in recent decades in computational models of polymer folding.  Folding models have been used to provide information about i) the initial collapse of polymer chains and the formation of transient structures~\cite{Shakhnovich1997, Dill1999, Yue1995}, ii) the behavior of chains of various lengths and morphologies~\cite{Wust2012, Farris2019, Wang2011}, iii) the effects of geometric constraints~\cite{Taylor2020Confine, Taylor2020Crowd, Ferreira2012}, iv) monomer site mutations~\cite{Shi2014}, and v) protein adsorption to surfaces~\cite{Radhakrishna2012}. Studies of single polymer chain folding have explored transitions from crystalline or globular conformations at low temperatures to extended or coil conformations at high temperatures. 

These transitions are sometimes correlated with maxima or minima in plots of the specific heat {\em versus} temperature~\cite{Wust2012, Farris2019, Farris2019Conf, Seaton2010}. However, since specific heat is an equilibrium property, it necessarily applies to quasi-equilibrium conditions (i.e., infinitesimally slow temperature changes), so specific heat is not likely to be a reliable predictor of conformation transitions during realistic heating and cooling rates.  In addition, folding kinetics are difficult to determine from computations because polymer chain relaxation times range between milliseconds and seconds. These times are beyond the accessible range of atomistic models such as molecular dynamics \cite{Dill1999, Privalov1996, Shakhnovich1997}.  Although Monte Carlo simulations can model folding kinetics~\cite{Ping2004, Gersappe1993, Castells2002, Cellmer2005, Zhange2008}, they are susceptible to becoming trapped in local metastable conformations and they are sensitive to the starting conditions. Consequently, it is usually necessary to average many Monte Carlo simulations to analyze even small conformation changes.

The steepest-entropy-ascent quantum thermodynamic (SEAQT) framework~\cite{Li2016a, Li2016b, Li2016c, Li2018, Li2018steepest, Li2018steepest, Li2017study, Li2018multiscale,mcdonald2021entropydriven, Yamada2018method, Yamada2019, Yamada2019kineticpartI, Yamada2019spin, Yamada2020kineticpartII,jhon2020,cano2015steepest,kusaba2019,kusaba2017,vonSpakovsky2020, Goswami2021,beretta2014steepest,beretta2006, Beretta1984, Beretta1985, Hatsopoulos1976-I, Hatsopoulos1976-IIa, Hatsopoulos1976-IIb, Hatsopoulos1976-III} is an alternative modeling strategy that can address these simulation challenges. For a given system, this framework predicts a thermodynamically unique kinetic path from any occupiable initial state to stable equilibrium. {\color{black} It utilizes a discrete energy spectrum generated by a Monte Carlo method, but folding kinetics are predicted separately in thermodynamic state space from a deterministic equation of motion --- not a Monte Carlo simulation. The kinetic path found by the SEAQT equation of motion proceeds from an initial state to a {\em global} stable state; it cannot remain stuck in metastable configurations nor is its path prediction subject to statistical uncertainties that would arise from predicting such a path from averaged Monte Carlo simulations. The sole statistical uncertainties that arise are those from the Monte Carlo method used to generate the energy spectrum.} 

The SEAQT equation of motion is derived from the principle of steepest-entropy-ascent, which has been suggested by Beretta \cite{Beretta2020} as a fourth law of thermodynamics.  This equation of motion predicts the non-equilibrium occupation probability distribution at every instance of time along the kinetic path. This leads to a straightforward description of the energy and entropy evolution in time. Moreover, tracking the change in the probability distribution makes it possible to calculate the non-equilibrium evolution of structural parameters such as radius of gyration, tortuosity, and end-to-end distance \cite{mcdonald2021entropydriven}.  The time evolution of such structural parameters provides an average physical representation of the chain at each point along the non-equilibrium kinetic path predicted by the equation of motion. Unlike properties commonly discussed in the literature, properties along the kinetic path are not limited by the usual quasi-equilibrium assumption.

In this contribution, we employ the Replica-Exchange Wang Landau algorithm to generate an energy landscape for a 58--monomer protein chain~\cite{Dill1993} and apply the {\color{black}SEAQT equation of motion to predict polymer chain folding along different non-equilibrium paths associated with either heating or cooling. These paths are then compared with quasi-equilibrium paths, which require no equation of motion and are simply a set of consecutive global stable equilibrium states. 
Once a given path is known, c}hain folding is characterized by the expected chain energy, radius of gyration, tortuosity, and end-to-end distance.

{\color{black}The remainder of the paper is organized into the following sections. In Section \ref{Method}, we begin with a description of the overall SEAQT framework. A non-Markovian Monte Carlo approach used to generate the energy landscape is presented as is the SEAQT equation of motion. This is followed by a description of how the state space results generated by the equation of motion are linked to the polymer chain conformations. Section \ref{Results} presents how different structural parameters (i.e., the radius of gyration, the tortuosity, and the end-to-end distance) evolve during cooling and heating along quasi-equilibrium and non-equilibrium paths. The evolution of polymer conformations are presented as well.  A discussion of these results is then given in Section \ref{Discussion} followed by a set of conclusions in Section \ref{Conclusions}.}

\section{Method}\label{Method}

{\color{black} Applying the SEAQT framework involves two essential steps. First, a discrete energy landscape is generated using an appropriate model: in this case, it is a lattice model~\cite{Lau1989}. The system is taken to be a single protein chain of fixed length, and each of the possible arrangements of the chain constitutes a conformation with a discrete energy, $E_j$. Borrowing terminology from quantum mechanics, we call $E_j$ an ``energy eigenlevel'' (or simply ``eigenlevel'' for short); the combination of a chain conformation and its corresponding energy eigenlevel is an ``eigenstate''. Many chain conformations (or configurations) have the same energy. The number of eigenstates an eigenlevel has is its degeneracy, $g$. The energy dependence of the degeneracy, $g(E_j)$, is sometimes called a density of states; here we refer to it as the energy degeneracy. The combination of the energy spectrum and associated degeneracies is referred to as the energy landscape.

Second, once the energy landscape is established, an initial state is chosen and the SEAQT equation of motion is solved over the landscape. This equation of motion provides the non-equilibrium path through thermodynamic state space from the initial state to that of global stable equilibrium (or stable equilibrium for short). The ``states'' in this context are thermodynamic states, which are not the same as eigenstates. Continuing with the quantum mechanical convention, the energy of a thermodynamic state is an expectation value calculated from the eigenlevel energies and their associated occupation probabilities. The time-dependence of the occupation probabilities are obtained from the SEAQT equation of motion.}

The details behind these two steps are provided in the subsequent sections. Section~\ref{SectionIIA} describes the Hamiltonian used to calculate the eigenlevels of a particular polymer chain, and Section~\ref{SectionIIB} describes the Wang-Landau algorithm used to construct the energy landscape. This is followed in Section~\ref{SectionIIC} by a description of the equation of motion. Section~\ref{SectionIID} then provides an explanation of how the predictions from the thermodynamic state space through which the SEAQT equation moves are linked to the expected conformations of the polymer chain.  

\subsection{Energy Landscape} \label{SectionIIA}
The eigenlevels that make up the energy landscape of a 58--monomer chain~\cite{Dill1993} are calculated with the hydrophobic-polar (HP) model formulated by Dill $et\;al.$~\cite{Lau1989}. In this 3-dimensional cubic lattice model, the polymer chain is composed of amino acids with either hydrophobic (nonpolar) or hydrophilic (polar) monomers. The energy of a particular chain conformation is calculated from the sum of the attractive interactions at hydrophobic contacts along folded segments of the chain. This energy, or Hamiltonian, is expressed by
 
{\color{black}
\begin{equation}
	E_j = \frac{1}{2}\, \underset {n = 1} {\overset {N} {\sum}} \, \sum_{\substack{m=1}}^{N_c} \varepsilon_{n, m}\;, \; 
	\left\{\!\begin{aligned}
	& \varepsilon_{n, m} = 0 \;\;\;\;\;\; \text{if \, $b_{n,m} = 0$} \\
	& \varepsilon_{n, m} = \varepsilon^{\text{\tiny HH}} \;\; \text{if \;  $b_{n,m} = 1$}
	\end{aligned}\right\}
	\label{PolyTotalEnergy}
\end{equation}     
where the quantity $\varepsilon_{n,m}$ represents an interaction energy between two monomers on nearest-neighbor sites of the cubic lattice. This term alternates between 0 and $\varepsilon^{\text{\tiny HH}}$ according to a Boolean switch, $b_{n,m}$, which is either 0 or 1 depending upon the identities of the $n$ and $m$ monomers. When $n$ and $m$ are (a) both polar monomers, (b) a polar-hydrophobic pair, or (c) consecutive monomers along the chain (i.e.~covalently bonded monomers), or (d) there is no monomer neighbor at $m$, then $b_{n,m}=0$ and $\varepsilon_{n, m}$ is set to 0. When $n$ and $m$ are two hydrophobic monomers that are not consecutive monomers along the chain (i.e., $n$ and $m$ are hydrophobic monomers brought into contact by chain folding), then $b_{n,m}=1$. In this circumstance,  $\varepsilon_{n, m}$ is set equal to the nonzero interaction energy, $\varepsilon^{\text{\tiny HH}}$. For the computations here,  $\varepsilon^{\text{\tiny HH}}$ is given the value $-1$ for convenience.  The first summation in Equation~(\ref{PolyTotalEnergy}) is over the $N=58$ monomers of the polymer chain while the second summation is over the $N_c$ first-nearest-neighbor sites to the $n^{th}$ monomer in the chain. For a cubic lattice, $N_c=6$. Only adjacent, non-bonded, hydrophobic monomers contribute to the Hamiltonian.} 

To generate the energy landscape of possible eigenstates (eigenenergies and conformations), a starting conformation is modified through a non-Markovian Monte Carlo procedure using the Wang-Landau algorithm~\cite{Wust2012, Farris2019}. The algorithm performs a random walk through possible conformations and maps both the system's spectrum of eigenenergies and their respective degeneracies. {\color{black} Following previous investigators~\cite{Deutsch1997, Wust2012, Farris2019}, allowable chain modifications in the Monte Carlo procedure include pull moves, reptation moves, rebridging moves, and pivot moves. The chosen distribution of movements is ${75\%}$ pull and reptation moves, ${23\%}$  rebridging moves, and ${2\%}$ pivot moves. These percentages were chosen by validating the degeneracy estimated with the Replica Exchange Wang-Landau algorithm against the exact degeneracy for the 14--monomer chain studied by Bachmann {\it et al.}~\cite{Bachmann2003}. 

Once the energy landscape is established, thermodynamic states are specified by the occupation probabilities of the energy eigenlevels at a given instant of time. These occupation probabilities can, for example, be predicted for a non-equilibrium path by the SEAQT equation of motion. Each thermodynamic state has energy $\langle E\, \rangle =\; \sum_j p_j \, E_{j}$ and entropy $\langle S\, \rangle =\; \sum_j p_j \,S_{j}$ where the angle brackets represent expectation values, $E_j$ and $S_j$ are the energy and entropy of the $j^{th}$ eigenlevel, respectively, and $p_j$ represents the occupation probability of eigenlevel $j$. If the set of all $p_j$'s satisfy a canonical distribution, then they represent a stable equilibrium state; if they do not, then they represent a non-equilibrium state.  }

\subsection{
Wang-Landau Algorithm}
\label{SectionIIB} 
{\color{black} The Wang-Landau algorithm employs entropic sampling to establish a system's discrete energy spectrum and degeneracies. The algorithm 
\cite{Vogel2013, Vogel2014, Vogel2014Conf} performs a random walk through the energy eigenlevels and estimates the degeneracies from the fact that the \textit{sampling} probability $p_{j}^s$ of the $E_j$ level approaches the inverse of the degeneracy provided the eigenlevels are sampled uniformly: $p_{j}^s \sim \frac{1}{g(E_j)}$. The random walk uses two counters. The first generates a histogram of the visits to each energy eigenlevel and the second provides a means for updating the degeneracy of the eigenlevels.  With each visit to a level, the histogram counter is incremented by 1 and the degeneracy is updated by a modification factor, $f$. Once all of the eigenlevels have been visited uniformly (a flat histogram), the modification factor is reduced, 
the histogram is reset, and the process is repeated until the modification factor reaches a small, predetermined threshold. Flatness is achieved when the minimum histogram value is greater than or equal to the average histogram value multiplied by a flatness criterion, which typically ranges between 80$\%$ and 100$\%$. In this work, the flatness criterion is set to ${99.2\%}$ based upon computational accuracy and efficiency considerations \cite{Wust2012}. The threshold criterion for the modification factor~\cite{Vogel2013,Vogel2014,Vogel2014Conf} is chosen as $\ln(f)<10^{-8}$.}

Although the Wang-Landau algorithm is applicable to any microcanonical ensemble, the characteristics of some physical systems slow the method's convergence. In polymer systems, the number of difficult-to-reach eigenlevels makes it hard to accurately estimate the degeneracy. This problem is mitigated by a parallelized version of the algorithm called the Replica Exchange Wang-Landau algorithm. Replica Exchange Wang-Landau divides the energy landscape into smaller energy ranges, or windows, with multiple Monte Carlo walkers moving independently in each energy window. To ensure convergence {\color{black}of} the degeneracy for the whole energy landscape, the energy windows partially overlap each other, and information is shared periodically among walkers from different energy windows \cite{Vogel2013, Vogel2014, Vogel2014Conf, Li2014}.
\begin{figure}[htbp]
\begin{center}
\includegraphics[width=.45\textwidth]{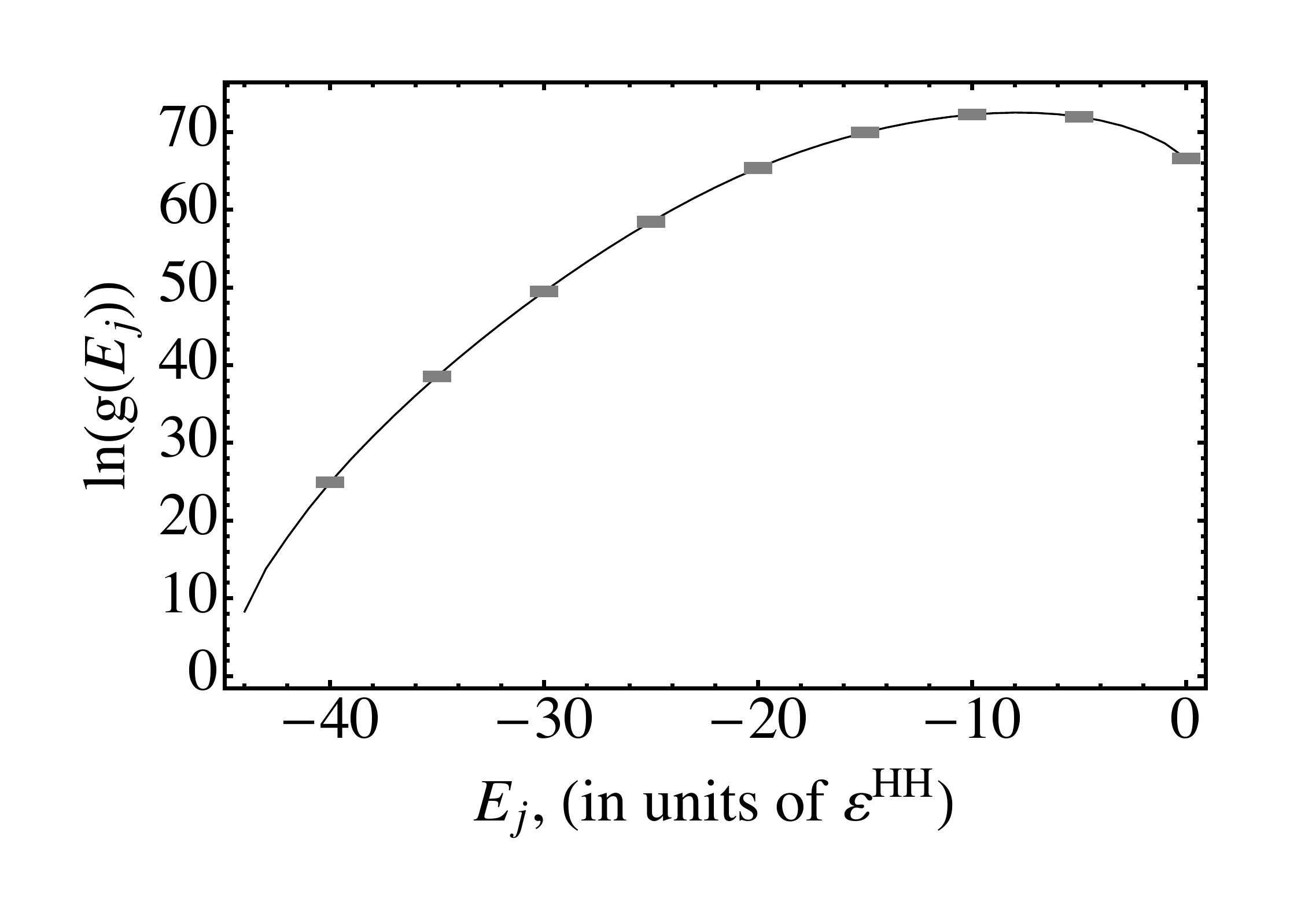}
\caption{The {\color{black}natural log of the degeneracy} of the $j^{th}$ energy eigenlevel versus the energy, $E_j$, of each level. The degeneracy is estimated for a 58--monomer polymer sequence calculated with the HP model using Replica Exchange Wang-Landau on a $60 \times 60 \times 60$ cubic lattice. The $E_j$ are summed interaction energies of the hydrophobic-hydrophobic bonds.}
\label{58DOS}
\end{center}
\end{figure}

\begin{figure}[htbp]
\begin{center}
\includegraphics[width=.45\textwidth]{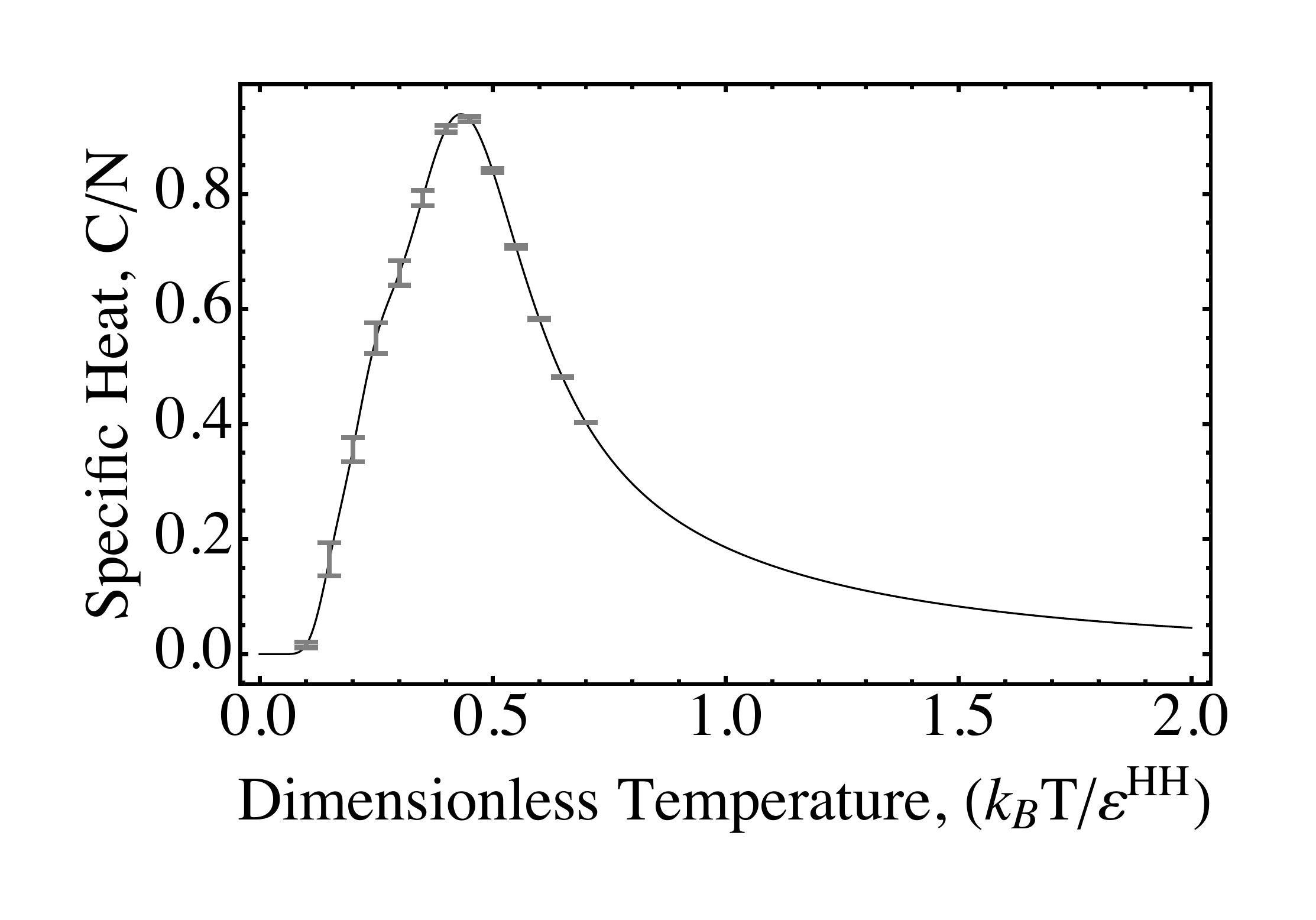}
\caption{The normalized specific heat plotted versus temperature for the 58--monomer chain.}
\label{58HC}
\end{center}
\end{figure}

Figure~\ref{58DOS} shows the degeneracy of the 58--monomer chain estimated with the Replica Exchange Wang-Landau algorithm. {\color{black}The algorithm precision is indicated at regular energy intervals by error-bars representing $\pm$ one standard deviation of the degeneracy determined from six independent simulations. The algorithm was implemented with three energy windows:  one from $0$ to $-35$ (units of $\varepsilon^{\text{\tiny HH}}$) containing one walker, a second from $-3$ to $-40$ containing one walker, and a third from $-6$ to $-44$ containing eight walkers. A large number of random walkers were used in the third energy window to accelerate the search for the lowest eigenlevel. Figure~\ref{58HC} represents the equilibrium specific heat calculated from the degeneracies of Figure~\ref{58DOS}. The error-bars are the upper and lower values of specific heat calculated from an `upper' and a `lower' degeneracy curve. These two degeneracy curves are obtained from the mean curve of Figure~\ref{58DOS} $\pm$ one standard deviation.}

\subsection{SEAQT Equation of Motion}\label{SectionIIC}

 The thermodynamic state of the polymer system is specified by a set of occupation probabilities associated with the energy eigenlevels of the system. These probabilities, the $p_j$, change with time as the system evolves from thermodynamic state to thermodynamic state. Given the system's time-independent energy landscape (its discrete energy spectrum and associated set of degeneracies established with the Replica-Exchange Wang-Landau algorithm, Figure~\ref{58DOS}), the equation of motion of the SEAQT framework{\color{black}, which is deterministic and not stochastic}, predicts the time evolution of a set of non-equilibrium probability distributions along a unique thermodynamic path starting from some arbitrary non-equilibrium state and ending in stable equilibrium. It does so without any \textit{a} \textit{priori} assumption of equilibrium. In other words, these non-equilibrium distributions are not equilibrium-derived. {\color{black}To the authors' knowledge, the Wang-Landau algorithm does not have any implicit dependence upon the equilibrium characteristics of the simulated system. In fact, the canonical distribution associated with equilibrium is unavailable until after the algorithm finishes estimating the energy landscape.}
 
 {\color{black}
 Another distinct feature of the equation of motion is that it requires no mechanistic detail about the system in order to calculate a kinetic path. The equation of motion can be applied to any energy landscape --- including a fictitious one with random eigenlevels and degeneracies --- because the equation of motion only depends upon the energy eigenlevels and their degeneracies. It is not affected by how the landscape is derived or even what physical system it describes. Thus, the SEAQT framework is able to predict a polymer chain's morphological evolution solely on the basis of the principle of steepest entropy ascent.} This principle reflects the redistribution of energy among available eigenlevels at every instant of time consistent with the laws of thermodynamics and mechanics. The steepest-entropy-ascent postulate in the framework is a variational principle that leads to the SEAQT equation of motion \cite{beretta2014steepest}. The question, of course, arises as to why the variational extremum should be a maximum rather than a minimum. 

It has been shown previously~\cite{Ziegler1957thermodynamik, Ziegler1963some, Ziegler1983chemical, Ziegler1983introduction, Ziegler1987principle} that both linear and nonlinear non-equilibrium thermodynamics can be deduced from Ziegler's maximum entropy production principle. This cannot be done with Prigogine's minimum entropy production principle \cite{Prigogine1967}, which is a special case of the Onsager-Gyarmati principle of linear non-equilibrium thermodynamics. Furthermore, much evidence suggests that processes in nature maximize entropy production at each instant of time~\cite{Yamada2020kineticpartII, Martyushev2013}, which is consistent with Ziegler's principle as well as the steepest-entropy-ascent principle of Beretta \cite{beretta2005generalPhD, Beretta2020,beretta2006, Beretta2009,Beretta1984, Beretta1985,beretta2014steepest}. In addition, as noted by Cahn and Mullins \cite{CahnMullins1964}, assuming entropy production is {\em minimized} fails in very simple cases such as that of ({\em{i}\/}) 1D steady state heat conduction and ({\em{ii}\/})  1D steady state mass diffusion in the presence of an externally maintained gradient. In contrast, as shown in Appendix B of Li, von Spakovsky, and Hin \cite{Li2018steepest} and by Li and von Spakovsky in \cite{Li2016b}, the SEAQT equation of motion leads to the correct steady state linear temperature and concentration profiles, respectively. This is done using the steepest-entropy-ascent principle only. No assumption is made of a particular kinetic mechanism, such as Fourier's law for transient thermal transport or Fick's law for transient mass transport~\cite{mcdonald2021entropydriven}.
           
Originally introduced and applied to small quantum systems in the 1970s and 1980s \cite{Beretta1984, Beretta1985, Hatsopoulos1976-I, Hatsopoulos1976-IIa, Hatsopoulos1976-IIb, Hatsopoulos1976-III}, the SEAQT framework has been extended over the past decade and a half to make it a practical tool for predicting non-equilibrium thermodynamic paths at all spatial and temporal scales. As a result, it has been applied to a wide variety of chemical and material problems \cite{Li2016a,Li2016b,Li2016c,Li2018,Li2018steepest,Li2018steepest,Li2017study, Li2018multiscale,mcdonald2021entropydriven,Yamada2018method,Yamada2019,Yamada2019kineticpartI,Yamada2019spin,Yamada2020kineticpartII,jhon2020,cano2015steepest,kusaba2019,kusaba2017,vonSpakovsky2020}.

The SEAQT equation of motion for a simple quantum system \cite{Beretta1984} (or a model which mimics a quantum system such as an Ising or a Potts model) is 
\begin{equation}
\color{black}
\frac {d \hat{\rho}} {dt}=\frac {1} {i \hbar}[\hat{\rho},\hat{H}]+\frac {1} {\tau(\hat{\rho})} {\hat D(\hat{\rho})} \label{EOM1}
\end{equation}
where $\hat{\rho}$ is the probability density operator representing the thermodynamic state of the system, $t$ is the time,  $\hbar$ is the modified Planck constant, and $\hat{H}$ is the Hamiltonian operator. If the system is classical, $\hat{\rho}$ becomes the set of occupation probabilities, $p_j$.  The left side of this equation and the first term on the right side constitutes the von Neumann form of the time-dependent part of the Schr\"odinger equation of motion. The state evolution of this part of the equation is strictly reversible. The final term on the right, in which $\tau$ is a relaxation parameter and $\hat{D}$ is a dissipation operator, is the SEAQT addition. This term adds the nonlinear dynamics of irreversible state evolution that are absent from the von Neumann or Schr\"odinger equation. Equation~(\ref{EOM1}) is valid for any irreversible process~\cite{beretta2014steepest,beretta2006, Beretta2009, Li2016a} and although originally simply postulated by Beretta $et \; al.$ \cite{Beretta1984}, it subsequently has been derived via a variational principle along the direction of steepest-entropy-ascent by a constrained gradient descent in Hilbert space while preserving the energy and occupation probabilities \cite{beretta2014steepest}.

Given the aforementioned HP model for the polymer chain (Section~\ref{SectionIIA}), the system is strictly classical (no quantum correlations). As a result, $\hat{\rho}$ and $\hat{H}$ commute since they are diagonal in the energy eigenvalue basis \cite{Li2016a,Li2016b,Li2018,beretta2006,Beretta2009,Li2016c}. For the case when the only generators of the motion (those operators which must be conserved) are the Hamiltonian and identity operator, the equation of motion can be expressed as
\begin{equation}
\color{black}
\frac{dp_j}{dt}=\frac {1} {\tau}\frac{\left|
\begin{array}{ccc}
 -p_j \ln \frac{p_j}{g_j} & p_j & {E}_j\, p_j \\
 \langle S \rangle & 1 & \langle E \rangle \\
 \langle ES \rangle & \langle E \rangle & \langle E^2 \rangle \\
\end{array}
\right|}{\left|
\begin{array}{cc}
 1 & \langle E \rangle \\
  \langle E \rangle & \langle E^2 \rangle \\
\end{array}
\right|}
\label{EOM2}
\end{equation}
where
\begin{eqnarray}
\color{black}
\langle E^2 \rangle &=& \underset {j} {{\sum}}\phantom{l}p_j\,{E}_j^2
\label{ExEn} \; , \nonumber \\
\color{black}
\langle E S \rangle &=& -\underset {j} {{\sum}}\phantom{l}p_j\,{E}_j \, \text{ln}\frac{p_j}{g_j} \; ,
\label{ExEnS}
\end {eqnarray}
and $p_j$, $E_j$, and $g_j$ are the occupation probability, energy, and degeneracy, respectively, of the $j^{th}$ energy eigenlevel. Quantities in angle brackets, $\langle \, \cdot \, \rangle$,  represent expectation values of properties such as the energy $\langle E \,\rangle$, entropy $\langle S \,\rangle$, and their product $\langle ES \,\rangle$. In this formulation, the von Neumann definition of entropy is used so that $S_j = -\ln \frac{p_j}{g_j}$. This form is chosen because it is the only one that satisfies all the characteristics required by the entropy of thermodynamics \cite{Gyftopoulos1997}.

Equation~(\ref{EOM2}) is only applicable to an isolated system, but it can be extended to account for multiple subsystems interacting within a larger isolated system \cite{Li2016a}. In doing so, mass and heat transfer between sets of subsystems can be modeled. For example, Equation~(\ref{EOM2}) can be modified to account for a heat interaction between two subsystems $A$ and $B$ so that the equation of motion of subsystem $A$ becomes~\cite{Li2016a}:
\begin{equation}
\color{black}
\frac{dp_j^A}{dt}=\frac {1} {\tau}\frac{\left|
\begin{array}{cccc}
 -p_j^A \text{ln}\frac{p_j^A}{g_j^A} & p_j^A &0 & {E}_j^A p_j^A \\
 \langle S^A \rangle & 1 & 0 &\langle E^A \rangle \\
 \langle S^B \rangle & 0 & 1 &\langle E^B \rangle \\
 \langle E\,S \rangle & \langle E^A \rangle & \langle E^B \rangle & \langle E^2 \rangle \\
\end{array}
\right|}{\left|
\begin{array}{ccc}
 1 & 0&
\langle E^A \rangle \\
  0&1&\langle E^B \rangle \\
   \langle E^A \rangle &\langle E^B \rangle &\langle E^2 \rangle
\end{array}
\right|} \;\; .
\label{ABEqM}
\end{equation}
Using the co-factors $C_1$, $C_2^A$, and $C_3$ determined from the first line of the determinant of the numerator and assuming the hypo-equilibrium condition developed by Li and von Spakovsky \cite{Li2016a}, this equation of motion for subsystem $A$ reduces to the following form \cite{Li2016a}:
\begin{align}
\frac {dp_j^A}{dt^*} & = p_j^A \left(-\text{ln}\frac{p_j^A}{g_j^A}-\frac{C_2^A}{C_1}-{E}_j^A\frac{C_3}{C_1}\right) \nonumber \\
&   = p_j\left[(S_j^A - \langle S^A \rangle)-({E}_j^A - \langle {E}^A\rangle)\frac{C_3}{C_1}\right] 
\label{AEqMCoF}
\end{align}
 with the variables $t$ and $\tau$ replaced by a dimensionless time
defined as $t^*= \int_0^t \frac{1}{\tau(\vec{p}(t'))}dt'$. The relaxation parameter, $\tau$, can either be assumed constant or taken to be a function of the time-dependent occupation probabilities, $p_j$, represented by the vector $\vec{p}$.  In what follows, $\tau$ is assumed to be a constant that scales the dimensionless time $t^*$ to real time. An equation of motion for subsystem $B$ is written in a similar fashion. However, if subsystem $B$ represents a large temperature reservoir, its state does not change in time so that its equation of motion is not needed, and the co-factor ratio in Equation~(\ref{AEqMCoF}) reduces to a function of the temperature of the reservoir, $T^R$, such that \cite{Li2016a, Li2016b}
\begin{align}
C_3/C_1= \frac{1}{k_{\text{\tiny B}} \,T^R} \equiv \beta^R \;\;\;\; .
\label{betaterm}
\end{align}

Now, in order to establish the initial condition used by the SEAQT equation of motion for the heating or cooling cases considered here, a procedure is needed to generate the occupation probabilities of this initial thermodynamic state.  Probabilities for an initial non-equilibrium thermodynamic state can be generated using a canonical distribution (for a different temperature), a partially canonical distribution, and a perturbation function.

The equation for the stable equilibrium probabilities, $p_j^{\text{se}}$, of a canonical distribution is given by
 	\begin{align}
 	p_j^{\text{se}}=\frac {\, g_j \exp(-\beta^{\text{se}} E_j)}{\underset {i} {{\sum}}\phantom{l}\, g_i \exp(-\beta^{\text{se}} \,E_i)}
 	\label{CanDist}
 	\end{align} 
where $\beta^{\text{se}}= 1/(k_{\text{\tiny B}} T^{\text{se}})$ and $T^{\text{se}}$ is the temperature of the canonical, i.e., stable equilibrium, state. For a chosen $T^{\text{se}}$, Equation~(\ref{CanDist}) provides all of the $p_j^{\text{se}}$. The expected energy for the stable equilibrium state with these occupation probabilities is
 	\begin{align}
	\langle E \, \rangle^{\text se} = \sum_{i}p_j^{\text{se}}\,E_i \;\;\;\; . \label{SysEse}
 	\end{align} 

For a partially canonical state, the equation for the partial equilibrium probabilities, $p_j^{\text{pe}}$, is written as
\begin{align}
p_j^{\text{pe}}=\frac {{\delta}_j \, g_j \exp(-\beta^{\text{pe}} \, E_j)}{\underset {i} {{\sum}}\phantom{l}{\delta}_i \, g_i \exp(-\beta^{\text{pe}} \, E_i)}
\label{PartCanDist}
\end{align}
where $\beta^{\textrm{pe}}$ has units of reciprocal energy and could be interpreted as being inversely proportional to a local temperature associated with the partially canonical state (metastable equilibrium state). The ${\delta}_j$ take values of 1 or 0 depending upon whether or not the energy eigenlevel is assumed to be occupied. Each set of 0 and 1 values defines a vector of values $\vec{\delta} = \{\delta_j, \; j=1,...,\}$ with the occupations of a particular partially canonical state. The expected energy for the partial equilibrium state with these occupation probabilities is
\begin{align}
\langle E \, \rangle^{\text pe} = \sum_{i}p_j^{\text{pe}} \,E_i \;\;\;\; . \label{SysEpe}
\end{align} 
For a chosen $\vec{\delta}$ and expected energy, Equations~(\ref{PartCanDist}) and (\ref{SysEpe}) are solved simultaneously for the $p_j^{\text{pe}}$ and $\beta^{\textrm{pe}}$. 

Once the $p_j^{se}$ and the $p_j^{pe}$ are known for a given system energy, $\langle E \rangle$, the following perturbation function can be used to generate a non-equilibrium probability distribution for time $t=t_0$:
\begin{equation} 
p_j^{\text{\tiny init}} = \lambda \; p_j^{\text{pe}}(E_j,\delta_j) + (1 - \lambda)\; p_j^{\text{se}}(E_j)  \;\; . \label{EOMintialDistribution}
\end{equation}	
Here, $\lambda$ is a number between 0 and 1. The closer it is to 1, the further the initial non-equilibrium state is from stable equilibrium. The non-equilibrium probability distribution of Equation~\ref{EOMintialDistribution} provides the thermodynamic state used for the initial condition of the SEAQT equation of motion, Equation~\ref{AEqMCoF}.

The properties and conformations resulting from the non-equilibrium distributions predicted by the SEAQT equation of motion along non-equilibrium paths are compared in Section \ref{Results} below to a corresponding set of properties and conformations resulting from quasi-equilibrium paths which consist only of a set of neighboring stable equilibrium states. The quasi-equilibrium paths and properties are based on the same system energy landscape used by the SEAQT equation of motion but are not predicted with the equation of motion; they consist only of stable equilibrium, i.e., canonical, states that can be generated using Equation~(\ref{CanDist}).

\subsection{Linking State Space to Chain Conformations}\label{SectionIID}

Previous Monte Carlo simulations of polymer chains provide quasi-equilibrium properties~\cite{Carmesin1988, Wust2012, Li2014}. The SEAQT equation of motion allows one to predict properties along any non-equilibrium thermodynamic path.  Its predictions are based exclusively on the way energy shifts among eigenlevels according to the steepest-entropy-ascent principle. The computational overhead for the equation of motion is very modest since it is a first-order ordinary differential equation in time. In addition, when the energy landscape is topographically rough, there is no danger of the system becoming trapped in a metastable state (i.e., a partially canonical state)~\cite{mcdonald2021entropydriven, Dill1999}, because the time derivatives of all the eigenlevel occupation probabilities do not become zero until the expectation values of the system energy and entropy --- the thermodynamic state --- satisfies the global stability criterion.

Nevertheless, in order to describe changes in chain conformations or changes in specific physical properties with time, it is necessary to work backward from the occupied eigenlevels and the expected values of thermodynamic states to representations (conformations) of the eigenstates. Because there are an astronomical number of chain conformations associated with each energy eigenlevel, the procedure outlined in Reference~\cite{mcdonald2021entropydriven} is used to link the thermodynamic states to expected conformations through microstructural descriptors. The descriptors are parameters calculated from a chain conformation. Those used here are the radius of gyration ($R_g$), the tortuosity ($\zeta$), 
and the end-to-end distance ($R_E$)~\cite{Carmesin1988,Wust2012,Li2014}. 

The radius of gyration is calculated via the expression   
\begin{equation}R_g = {\left(\frac{1}{N} \sum_i^N   \lVert \boldsymbol{r}_i-\boldsymbol{r}_{\textrm{cm}}\rVert{}^2\right)}^{\frac{1}{2}}
\end{equation}
where $N$ is the number of monomers in the system, $\boldsymbol{r}_i$ is the 3-dimensional location vector of a given monomer, and $\boldsymbol{r}_{\textrm{cm}}$ is the calculated location of the center of mass of the chain. All the monomers are assumed to have the same mass. The radius of gyration provides a quantitative assessment of the distribution of mass of the polymer chain.

The tortuosity employed here is defined as
\begin{equation}
\zeta ={\left(\frac{1}{N-2} \sum _i^{N-2} \lVert\boldsymbol{s}_i-\bar{\boldsymbol{s}}\rVert{}^2\right)}^{\frac{1}{2}}
\end{equation}
 where $\boldsymbol{s}_i$ is a local vector at the $i^{th}$ monomer that reflects the cumulative turns along the chain and $\bar{\boldsymbol{s}}$ is the mean of these vectors. The vector $\boldsymbol{s}_i$ has components
\begin{equation*}
\boldsymbol{s}_i= \left( \begin{array}{cc}
   \sum_{j=1}^i \, w_{j_{x}} \\[2mm]
   \sum_{j=1}^i \, w_{j_{y}} \\[2mm]
   \sum_{j=1}^i \, w_{j_{z}} \\
  \end{array}  \right)
  \end{equation*}
 with the ``turn vector,'' $\boldsymbol{w}_j$, given by
  \begin{equation*}
\boldsymbol{w}_j=\boldsymbol{r}_{j,j+1}\times \boldsymbol{r}_{j,j+2} , \;\;\;\;\; 1 \leq i \leq (N-2)  \;\; .
\end{equation*}
To calculate $\boldsymbol{s}_i$, the cross product of vectors from the current monomer to the next two monomers in the sequence (i.e., $\boldsymbol{r}_{j,j+1}=\boldsymbol{r}_{j+1}-\boldsymbol{r}_{j}$ and $\boldsymbol{r}_{j,j+2}=\boldsymbol{r}_{j+2}-\boldsymbol{r}_{j}$) are determined and their {\it x, y}, and {\it z} components summed. Whenever the polymer chain bends, it produces a non-zero $\boldsymbol{w}_j$ that is oriented perpendicular to the plane of the turn. The tortuosity is determined from the difference between the cumulative mean turns along the entire length of the chain, and it represents a measure proportional to the number of turns present along the chain segments. It is zero for a straight chain.

The end-to-end distance, $R_E$, given by
\begin{eqnarray}
R_E &=& \lVert\boldsymbol{r}_0-\boldsymbol{r}_N\rVert  \\[2mm]
&=& 
\sqrt{(x_0-x_N)^2 + (y_0-y_N)^2 + (z_0-z_N)^2}  \nonumber
\end{eqnarray}
is the distance from the head of the chain, with coordinates $\boldsymbol{r}_0 = \{x_0,y_0,z_0\}$, to the tail of the chain, with coordinates $\boldsymbol{r}_N = \{x_N,y_N,z_N\}$.

 A direct link between state space and chain conformations would involve associating the structural parameters with the conformations in each energy eigenlevel, but this approach is impractical because the degeneracy of each $E_j$ is far too large to record but some small fraction of the chain conformations~\cite{mcdonald2021entropydriven, Bachmann2003}. Instead, a strategy is employed to identify and record only those conformations that lie along the kinetic path. First, at each energy eigenlevel sampled by the Replica Exchange Wang-Landau algorithm, {\color{black} the arithmetic average of the descriptors of unique visits is found for $R_g$, $\tau$, and  $R_E$. These averages are determined from over $10^{11}$ unique visits to each eigenlevel (degeneracies are in the range $10^{8}<g<10^{30}$).  Wust {\em et al.~}\cite{Wust2012} also used this re-sampling technique and noted that it ``cover[s] conformational space more uniformly and faster (including conformational regions of low energy) than with standard multicanonical sampling.'' 

Next, the SEAQT equation of motion is solved for the occupation probabilities of the energy eigenlevels along a given kinetic path. Once these predicted probabilities are known, they are multiplied by the arithmetic averages of $R_g$, $\tau$, and  $R_E$ at each energy eigenlevel and summed to obtain expectation values of these descriptors along the kinetic path. Finally, to determine representative chain conformations along the path, the Replica Exchange Wang-Landau algorithm is run a second time to record specific conformations.  During this second Monte Carlo simulation, conformations are recorded only for occupied eigenlevels and descriptor values close to the expected values obtained from the SEAQT equation of motion. Since the number of occupied eigenlevels is much less than the total eigenlevels, and matching expected descriptor values eliminates all but a few of the degenerate conformations, the procedure yields a small set of conformations that evolve smoothly with time along the kinetic path.}

Scaling the rate with which the simulated system traverses the kinetic path to real-time can be accomplished by linking $\tau$ to physical quantities like a diffusion coefficient or other experimental data. How this is done is explained in Section \ref{Discussion}.

\section{Results}\label{Results}

\begin{figure}[htbp]
           \begin{center}
                       \includegraphics[width=0.45\textwidth]{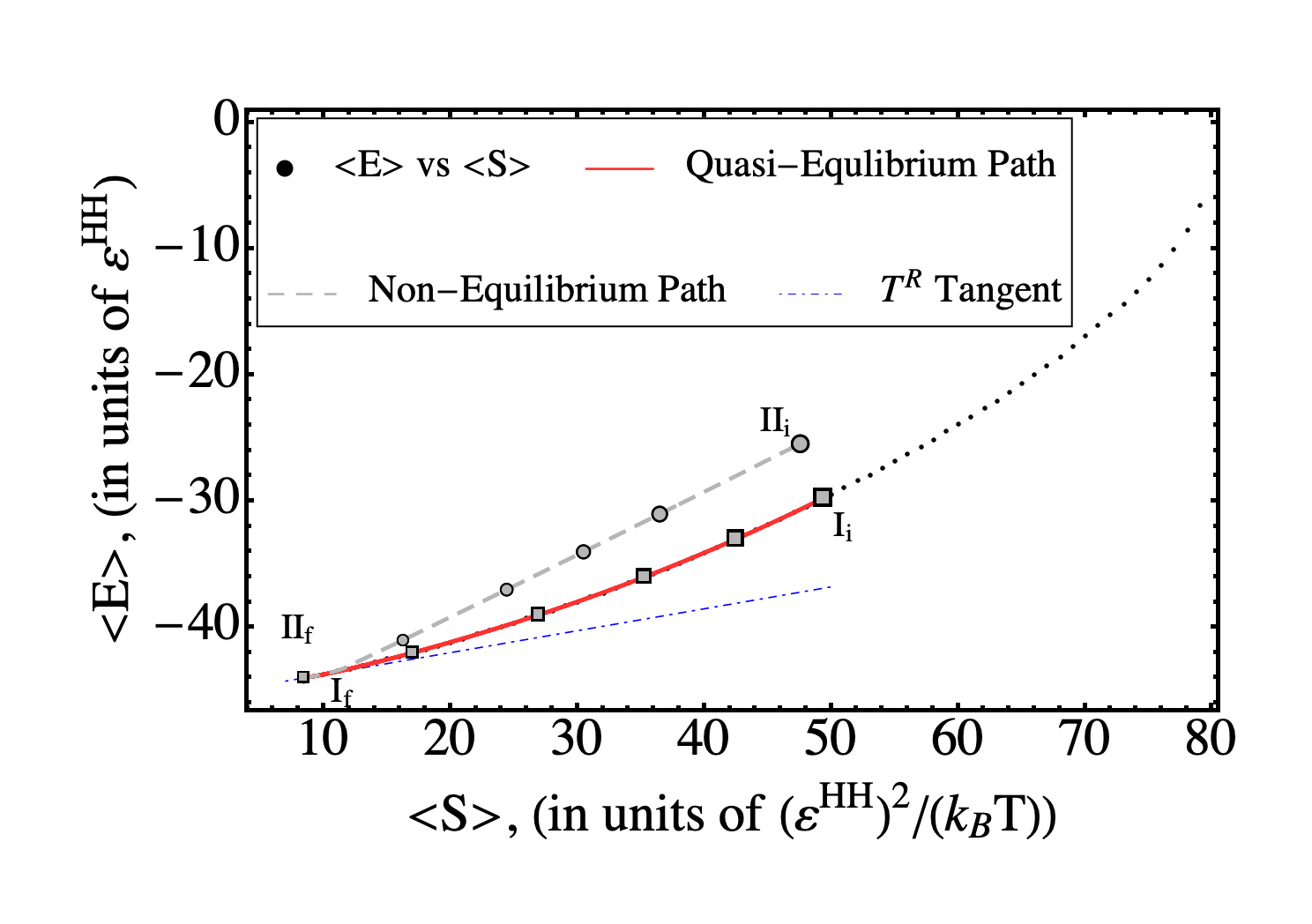}
                       \caption{The energy versus entropy quasi-equilibrium path (solid curve labeled I, red online) and
                       non-equilibrium path (dashed curve labeled II) associated with the 58--monomer chain system during cooling. The circles and squares correspond to the energy and entropy for the chain conformations shown in Figure~\ref{Low_T_Chains}. The largest circle and square are the initial states of the two paths, marked I$_i$ and II$_i$, while the smallest circle and square are the final stable equilibrium state at the reservoir temperature, marked I$_f$ and II$_f$. The quasi-equilibrium line plus the dotted curve that extrapolates beyond it form a curve of equilibrium states; the slope of this curve at any point represents the temperature and the dot-dashed slope line at the lower left is the reservoir temperature.}
                       \label{Low_T_Path}
           \end{center}
\end{figure}

\begin{figure}[htbp]
           \begin{center}
                       \includegraphics[width=0.45\textwidth]{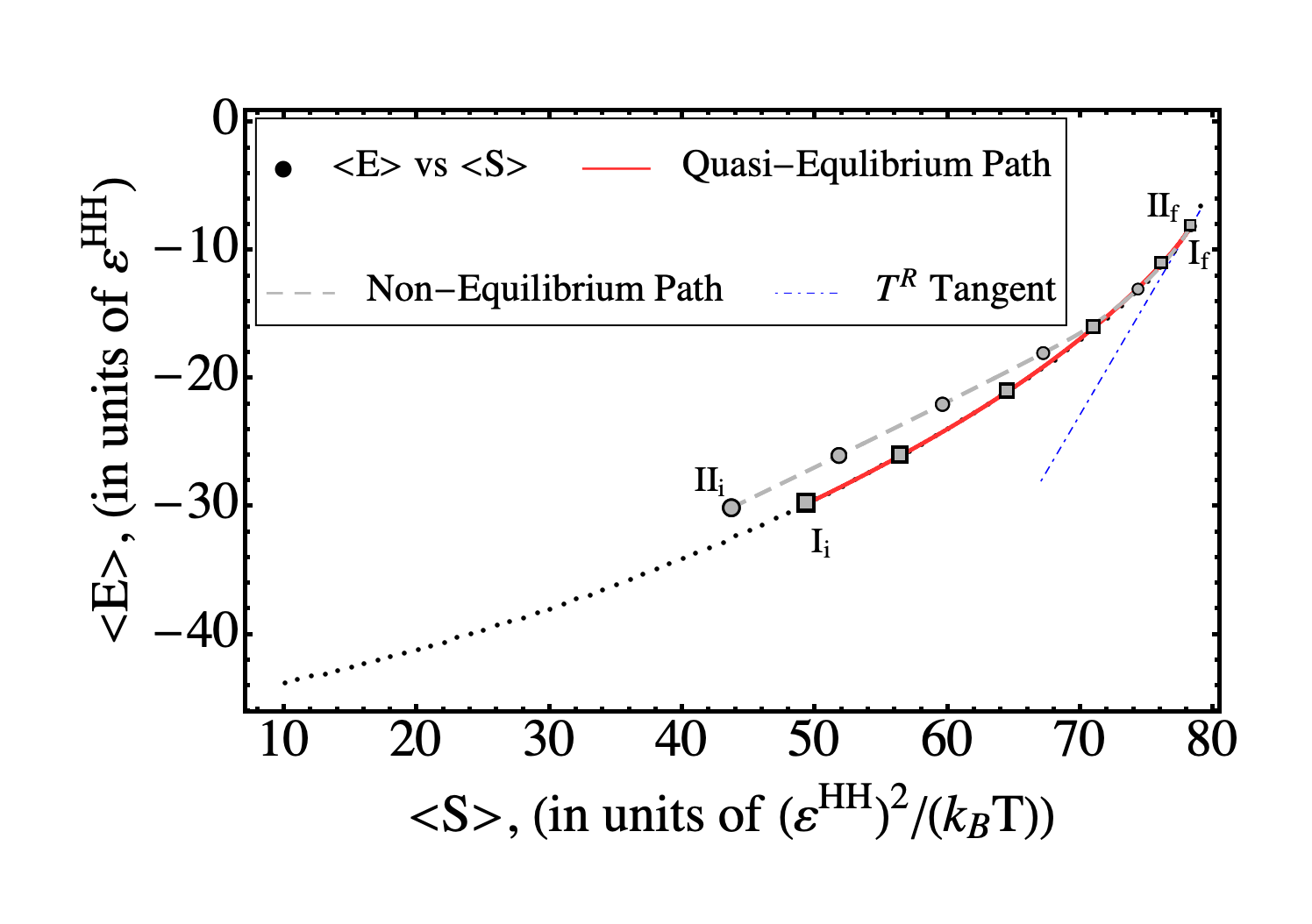}
                       \caption{The energy versus entropy quasi-equilibrium path (solid curve labeled I, red online) and
                       non-equilibrium path (dashed curve labeled II) associated with the 58--monomer chain system during heating. The circles and squares correspond to the energy and entropy for the chain conformations shown in Figure~\ref{High_T_Chains}. The largest circle and square are the initial states of the two paths, marked I$_i$ and II$_i$, while the smallest circle and square are the final stable equilibrium state at the reservoir temperature, marked I$_f$ and II$_f$. The quasi-equilibrium line plus the dotted curve that extrapolates beyond it form a curve of equilibrium states; the slope at any point along this curve represents the temperature and the dot-dashed slope line at the upper right is the reservoir temperature.}
                       \label{High_T_Path}
           \end{center}
\end{figure}

The evolution of the 58--monomer chain is tracked for two types of thermal interactions: an exchange of heat with a low-temperature reservoir (cooling), and an exchange of heat with a high-temperature reservoir (heating). For each of these cases, two kinetic paths are {\color{black}considered: (I) a quasi-equilibrium path and (II) a non-equilibrium path. The latter evolves according to the equation of motion from some initial non-equilibrium thermodynamic state to stable equilibrium with the thermal reservoir.  The quasi-equilibrium path is an idealized reversible path for which equilibrium is maintained continually until thermal equilibrium with the thermal reservoir is achieved. The equation of motion is not needed to describe this path because occupation probabilities obey a canonical distribution (Equation~(\ref{CanDist})) at each temperature along the path. For the non-equilibrium path, however, the initial non-equilibrium thermodynamic state is established using Equations~(\ref{CanDist}) through (\ref{EOMintialDistribution}) and the final stable equilibrium state is set by the choice of the reservoir temperature. The occupation probabilities at each instant of time along the path between these two states must be found from the equation of motion. The final thermodynamic state predicted by the equation of motion coincides with the equilibrium canonical distribution at the reservoir temperature. 

Results for the quasi-equilibrium path (I) and non-equilibrium path (II) for cooling are shown in Figures~\ref{Low_T_Path} to \ref{Low_T_Chains} and for heating in Figures~\ref{High_T_Path} to \ref{High_T_Chains}. The thermodynamic initial and final states are indicated by subscripts ``$i\/$'' and ``$f\/$'', respectively. For the two cooling paths seen in Figure~\ref{Low_T_Path}, the initial thermodynamic states for the quasi-equilibrium and non-equilibrium paths are different: $\text{I}_i \neq \text{II}_i$. The low-temperature reservoir is set to the same temperature for both paths so that their final states, I$_f$ and  II$_f$, are the same stable equilibrium state. For the heating processes of Figure~\ref{High_T_Path}, the initial states for the quasi-equilibrium and non-equilibrium paths have the same expected energy but different probability distributions: $\text{I}_i \neq \text{II}_i$), so they are different thermodynamic states. The high-temperature reservoir is set to the same temperature for both paths, so again, their final states, I$_f$ and  II$_f$, are the same.

In both Figures~\ref{Low_T_Path} and \ref{High_T_Path}, the solid and dotted curves together represent the curve of stable equilibria; the slope of this curve is the thermodynamic temperature, $(\frac{\partial{E}}{\partial{S}})_{\textit{\tiny V,N}}\,$. The $\langle E\,\rangle \, vs \, \langle S\,\rangle$ curve of stable equilibria can be calculated from the canonical probability distribution over a range of temperatures, $T^{se}$.  The solid curve (red online) represents the quasi-equilibrium path chosen for the cooling case. The dashed gray curve is the non-equilibrium path predicted by the SEAQT equation of motion.  For both paths, the black circles and squares are states at selected times along the paths from the initial state to the final equilibrium state.  During cooling along either path in Figure~\ref{Low_T_Path}, the energy and entropy of the polymer decrease as energy is extracted from the polymer and dumped into the reservoir. During heating (see Figure~\ref{High_T_Path}), the energy and entropy of the polymer increases as energy flows from the reservoir into the polymer. 

To compute a single non-equilibrium path, the SEAQT equation of motion (a system of first-order ordinary differential equations in time) is solved using Matlab's ODE45 numerical differential equation solver. Solving the system requires about a minute of CPU time on a processing thread of a Mac computer using an Intel i7 processor 8750H clocked at 2.2 GHz. The solution speed can be adjusted by varying the mass matrix through Matlab's {\em odeset} options with the parameters {\em RelTol} and {\em AbsTol} set to $1^{-10}$ and $1^{-20}$ respectively. 
Of course, the total computational time to calculate SEAQT kinetics must also include the time needed to build the energy landscape.  This is the most resource-intensive step in the calculations. For the degeneracy shown in Figure~\ref{58DOS}, the arithmetic averages of the descriptors, and representative chain conformations required approximately one week of CPU time, although the additional simulations needed to generate the error-bars extended the computational time to about one month of CPU time.}

\subsection{Interactions with a Low-T Reservoir}\label{low-T-Results}

The final states, I$_f$ and II$_f$, in Figure~\ref{Low_T_Path} correspond to a low-temperature reservoir at a non-dimensional temperature of $k_BT^R/\varepsilon^{\text{\tiny HH}} = 0.125$ where $\varepsilon^{\text{\tiny HH}}$ is the negative of the interaction energy in Equation~(\ref{PolyTotalEnergy}) and $k_B=1$ \cite{Wust2012}. The two paths through state space in Figure~\ref{Low_T_Path} indicate the difference between an {\em ideal, reversible} path, i.e., the quasi-equilibrium path, and a {\em real} path, i.e., the non-equilibrium path, during cooling. Next, we explore how the parameters associated with the chain descriptors and conformation evolve along these two kinetic paths.

\begin{figure}[htbp]
           \begin{center}
                       \includegraphics[width=0.45\textwidth]{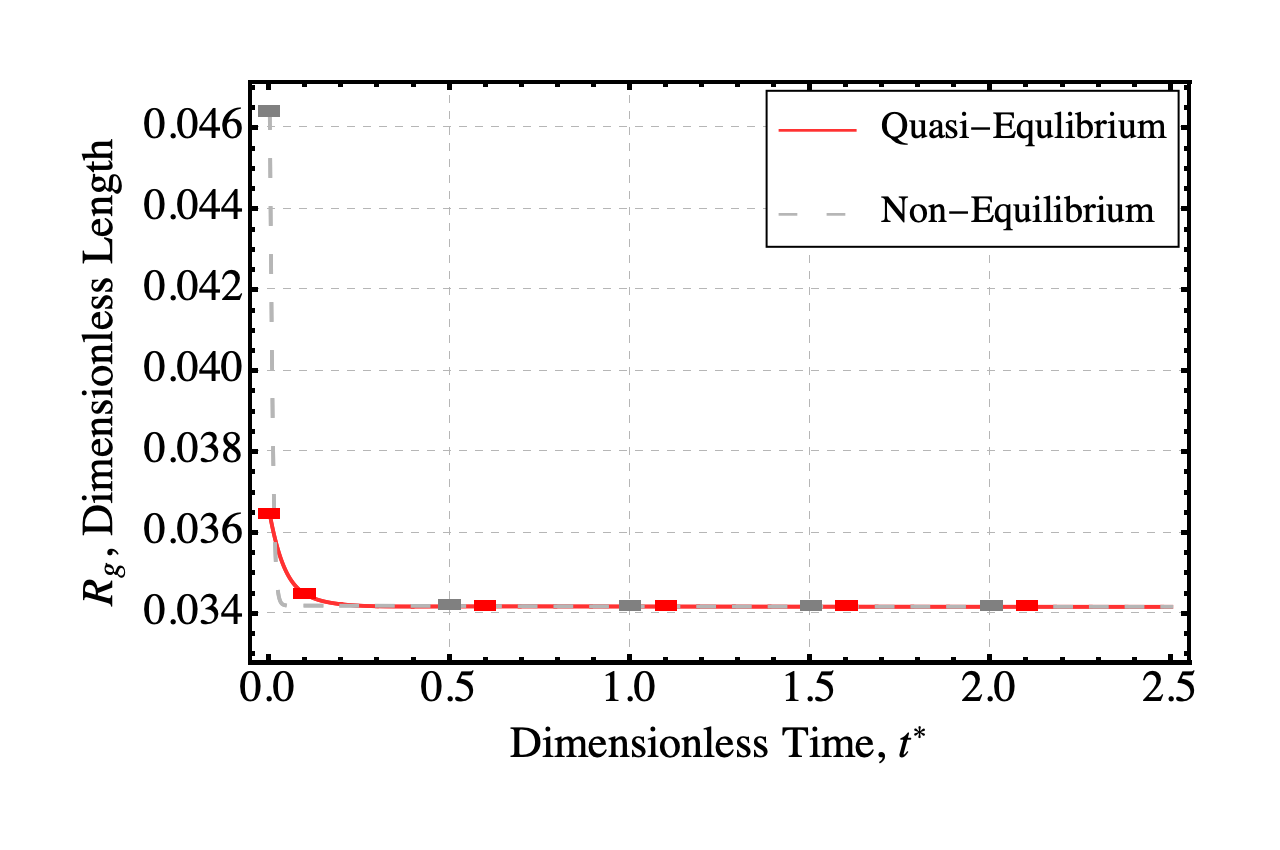}
                       \vspace{0.5truecm}
                       \includegraphics[width=0.45\textwidth]{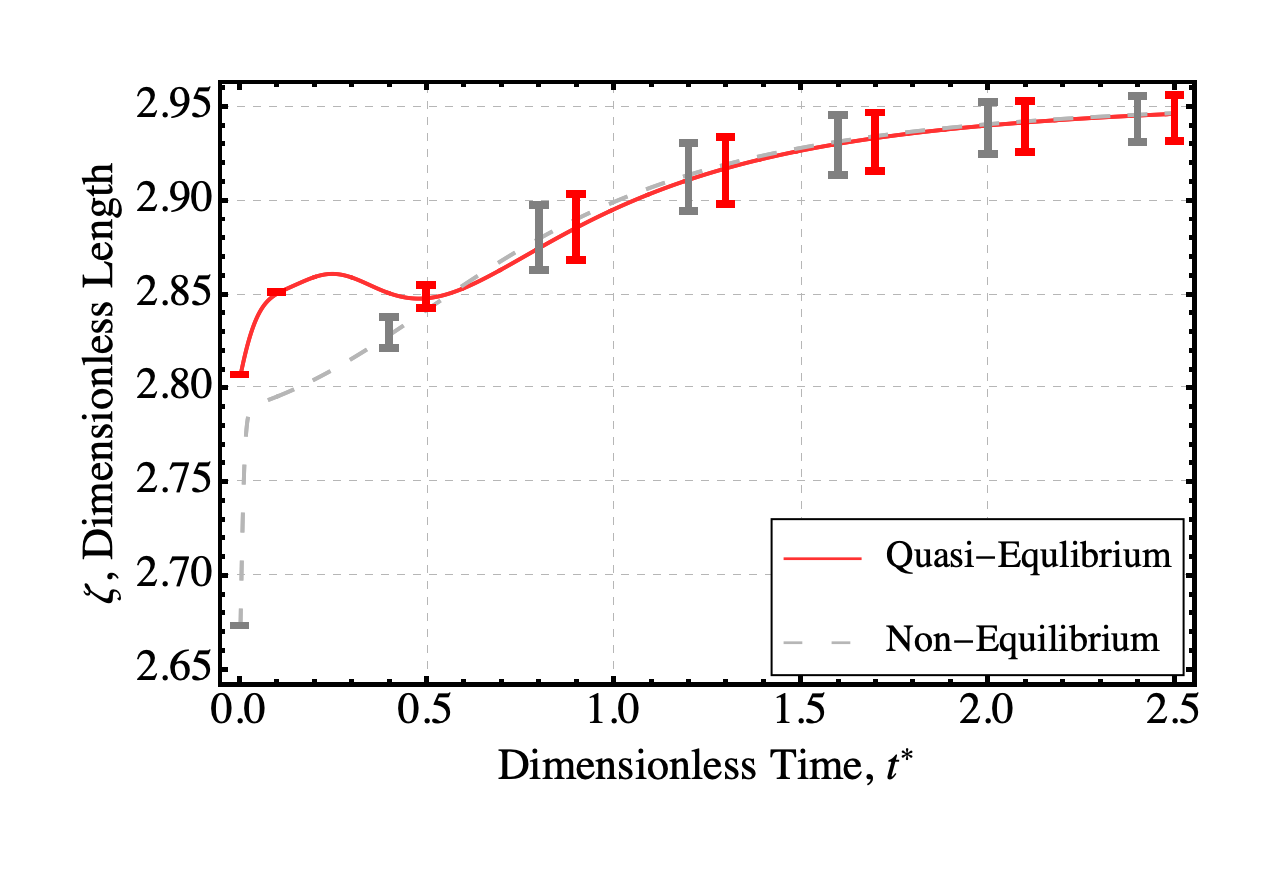}
                       \vspace{0.5truecm}
                       \includegraphics[width=0.45\textwidth]{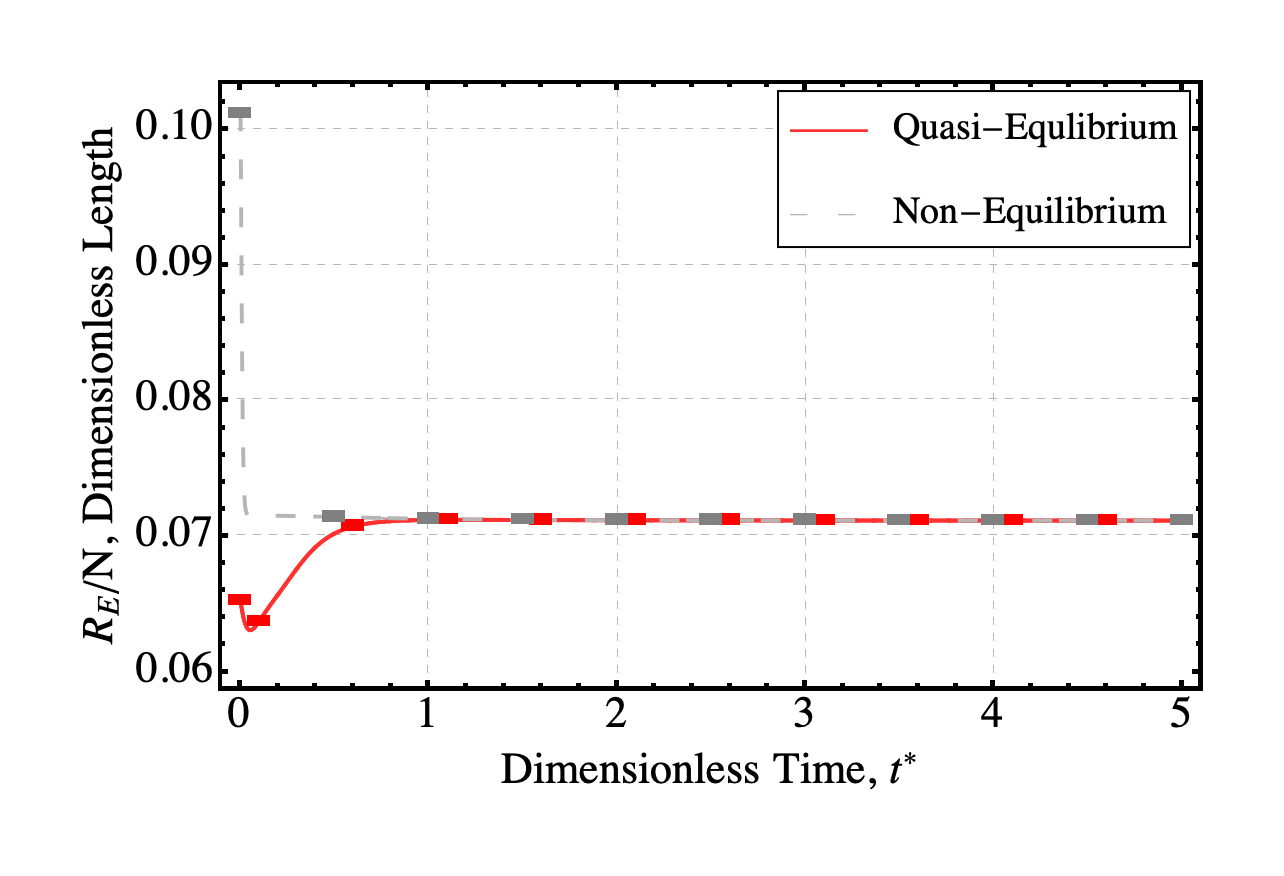}
                       \caption{Evolution of the expectation values of the radius of gyration, $R_g$, the tortuosity, $\zeta$, and the end-to-end distance, $R_E$, along the quasi-equilibrium (solid) and non-equilibrium (dashed) cooling paths.}
                       \label{Low_Struct_Param}
           \end{center}
\end{figure}

{\color{black} The  time evolution of the descriptors $R_g$, $\tau$, and  $R_E$ are shown in Figure~\ref{Low_Struct_Param}. The SEAQT equation of motion is a deterministic equation that does not introduce error into the calculation of descriptor properties, so the error-bars in Figure~\ref{Low_Struct_Param} reflect the range of values expected from uncertainty in the degeneracies of Figure~\ref{58DOS}.  The upper error-bars in Figure~\ref{Low_Struct_Param} were calculated using the upper limit of the degeneracies in Figure~\ref{58DOS} (i.e., $\ln\,g(E_j) + \sigma$) and the lower error-bars in Figure~\ref{Low_Struct_Param} were calculated using the lower limit of the degeneracies ($\ln\,g(E_j) - \sigma$). } 

The expectation values for the radius of gyration, $R_g$, nearly overlap during quasi-equilibrium cooling and non-equilibrium cooling. Although the curves are similar, the initial value of $R_g$ for the non-equilibrium curve is significantly larger, suggesting it has a less-pronounced hydrophobic core and, thus, a wider distribution of mass around its core than the initial conformation for the quasi-equilibrium path. The expectation values for $R_g$ along both paths decrease rapidly at the beginning of the evolutions and then very gradually as the stable equilibrium value at the reservoir temperature is approached.

For the tortuosity descriptor, there is an initial rapid increase in $\tau$ during non-equilibrium cooling followed by a more gradual change to the final equilibrium value.  There is a somewhat less rapid initial change followed by more moderate evolution along the quasi-equilibrium path, although the total change in tortuosity along the quasi-equilibrium path is about half the change along the non-equilibrium path. {\color{black} It is not clear why initially the quasi-equilibrium tortuosity exhibits a non-monotonic increase then decrease, but similar changes in the slope of tortuosity versus temperature curves have been reported in other studies~\cite{Wust2012}.} 

The end-to-end distance, $R_E$, during cooling follows a trend similar to the radius of gyration for the non-equilibrium path. For the quasi-equilibrium path, there is a small initial rapid decrease followed by a gradual increase to the stable equilibrium value. The difference between the $R_E$ values of non-equilibrium and quasi-equilibrium cooling is initially large but sharply decreases with time. For both $R_E$ paths, {\color{black} there is little change after  $t^* \sim 0.5$ at which time $R_E$ essentially has reached the stable equilibrium value at the low-temperature reservoir.}

\begin{figure*}[htbp]
           \begin{center}
                       \includegraphics[width=0.95\textwidth]{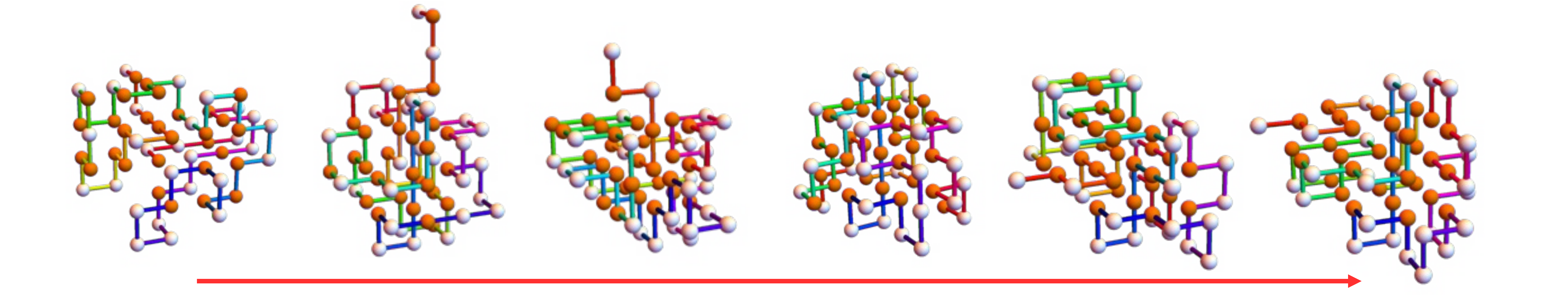}\\[5mm]
                       \includegraphics[width=0.95\textwidth]{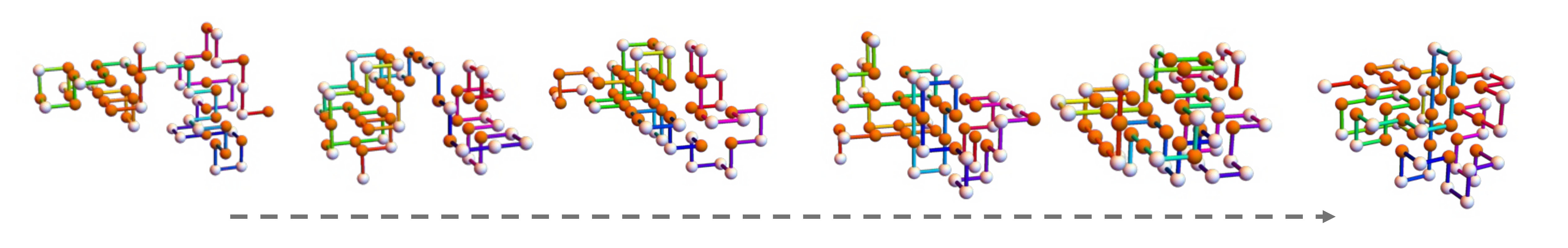}
                       \caption{Polymer conformations along quasi-equilibrium (top) and non-equilibrium (bottom) cooling paths. The six configurations of each path correspond to the states indicated by squares/circles in Figure~\ref{Low_T_Path}. The arrows indicate the direction of evolution from the initial conformation (left side) to the equilibrium conformation at the reservoir temperature (right side).  Both paths end at the same equilibrium state. Darker spheres (orange online) represent hydrophobic monomers; lighter spheres (white online) represent polar molecules. }
                       \label{Low_T_Chains}
           \end{center}
\end{figure*}

The expected chain conformations along the non-equilibrium and quasi-equilibrium paths of Figure~\ref{Low_T_Path} are presented in Figure~\ref{Low_T_Chains}. The upper row shows six expected conformations along quasi-equilibrium path I in Figure~\ref{Low_T_Path} at the points indicated by squares in that figure. The solid arrow in Figure~\ref{Low_T_Chains} indicates the direction of evolution in time from a representative initial configuration on the left to stable equilibrium on the right. The lower row shows six expected conformations along non-equilibrium path II in Figure~\ref{Low_T_Path} at the points indicated by circles in that figure. The dashed arrow in Figure~\ref{Low_T_Chains} indicates the direction of evolution in time from a representative initial configuration on the left to stable equilibrium on the right.  These conformations are chosen from {\em expected values} for the energy and the structural descriptors, $R_g$, $\tau$, and $R_E$, for individual states along the calculated paths. The conformations shown are not simply snapshots taken from a continuous Monte Carlo simulation, but rather are structures that represent expected conformations of the chosen thermodynamic states along the calculated path. For the quasi-equilibrium path, the initial chain conformation (left-most image) has a well-formed hydrophobic core (the darker hued orange spheres represent the hydrophobic monomers) with only a few free monomers apart from the center of mass.  The initial state for the non-equilibrium path has a smaller hydrophobic core and segments of non-interacting monomers extending away from the core are more apparent.  The two paths have different initial states, but the right-most image on both rows are same because the final stable equilibrium state for the quasi-equilibrium and non-equilibrium paths are identical. Generally speaking, the quasi-equilibrium conformations begin with a relatively compact core that becomes slightly denser as the chain cools to the reservoir temperature, whereas the non-equilibrium path begins with a more open conformation that eventually collapses to the same compact core.

As eigenlevel occupation evolves differently with time along the quasi-equilibrium and non-equilibrium paths, chain conformations must also evolve differently. Conformation changes along the two kinds of paths are wholly distinct even though the paths have similar descriptor values because the values do not all change at the same speed. Since only expected descriptor values are considered in this state-based method, the claim made here is not that an actual system will adopt the exact predicted chain conformations, but instead that the chains will appear notably similar. Moreover, despite the fact that Wang-Landau chain movements --- such as re-bridging moves --- are random and unrelated to the sequence of changes in a particular chain, the SEAQT framework can predict chain evolution when a continuous set of descriptors are selected along the kinetic path to ensure conformations evolve smoothly.

 \subsection{Interactions with a High-T Reservoir}\label{high-T-Results}

To investigate the behavior of the 58--monomer chain during a heating process, the SEAQT equation of motion is solved with a higher reservoir temperature. In this case, heat flows from the reservoir into the polymer until the system and high-temperature reservoir equilibrate. The final states, I$_f$ and II$_f$, in Figure~\ref{High_T_Path} correspond to a high-temperature reservoir at a non-dimensional temperature of $k_BT^R/\varepsilon^{\text{\tiny HH}} = 1.8$. The quasi-equilibrium (I) and non-equilibrium (II) heating paths are shown in Figure~\ref{High_T_Path}. During heating,  the polymer chain is expected to transition from a globular conformation at its initial state to an uncoiled conformation at the high reservoir temperature.  Because the degeneracy of the eigenlevels greatly increases at higher energies (Figure~\ref{58DOS}), the difference between the quasi- and non-equilibrium paths is also expected to be more pronounced for the heating case.

\begin{figure}[htbp]
           \begin{center}
                       \includegraphics[width=0.45\textwidth]{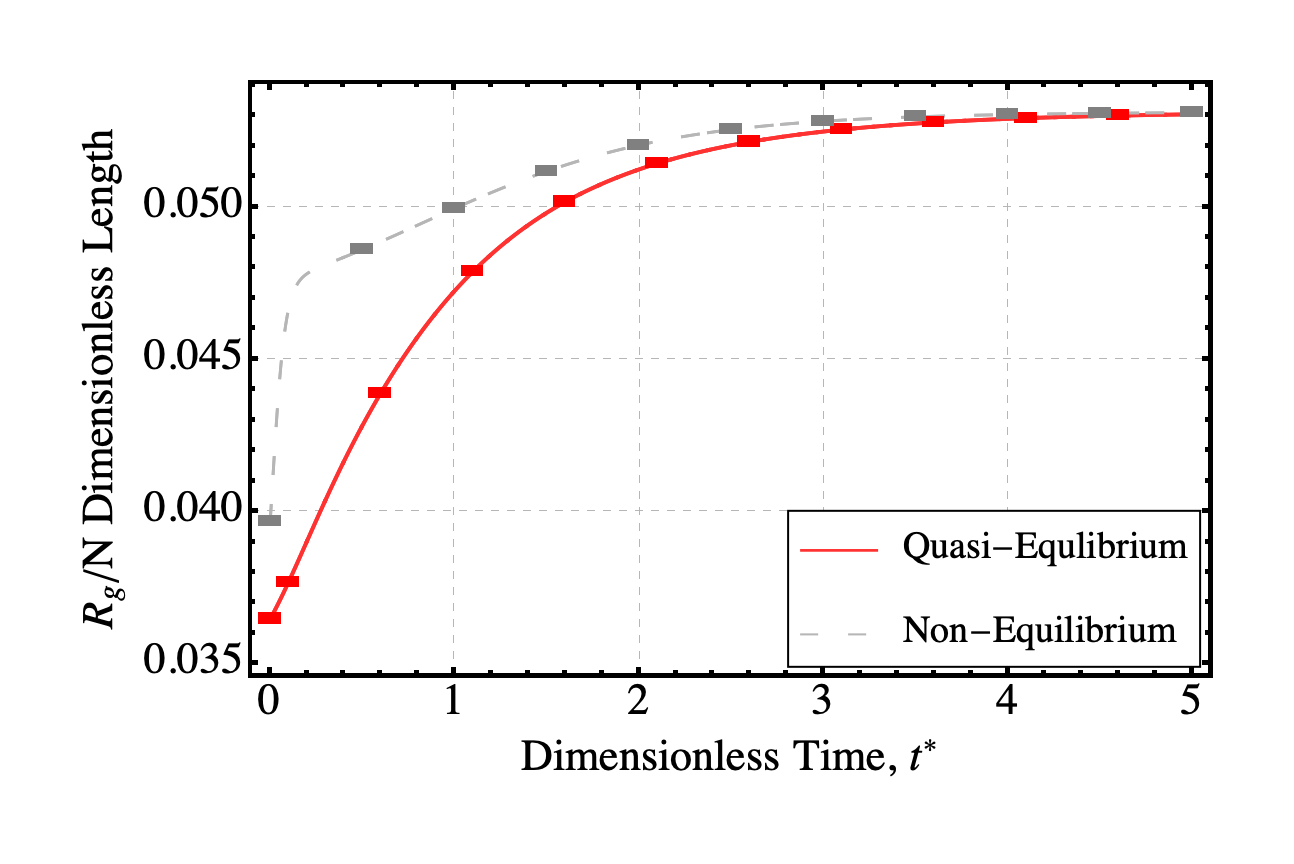}
                       \vspace{0.5truecm}
                       \includegraphics[width=0.45\textwidth]{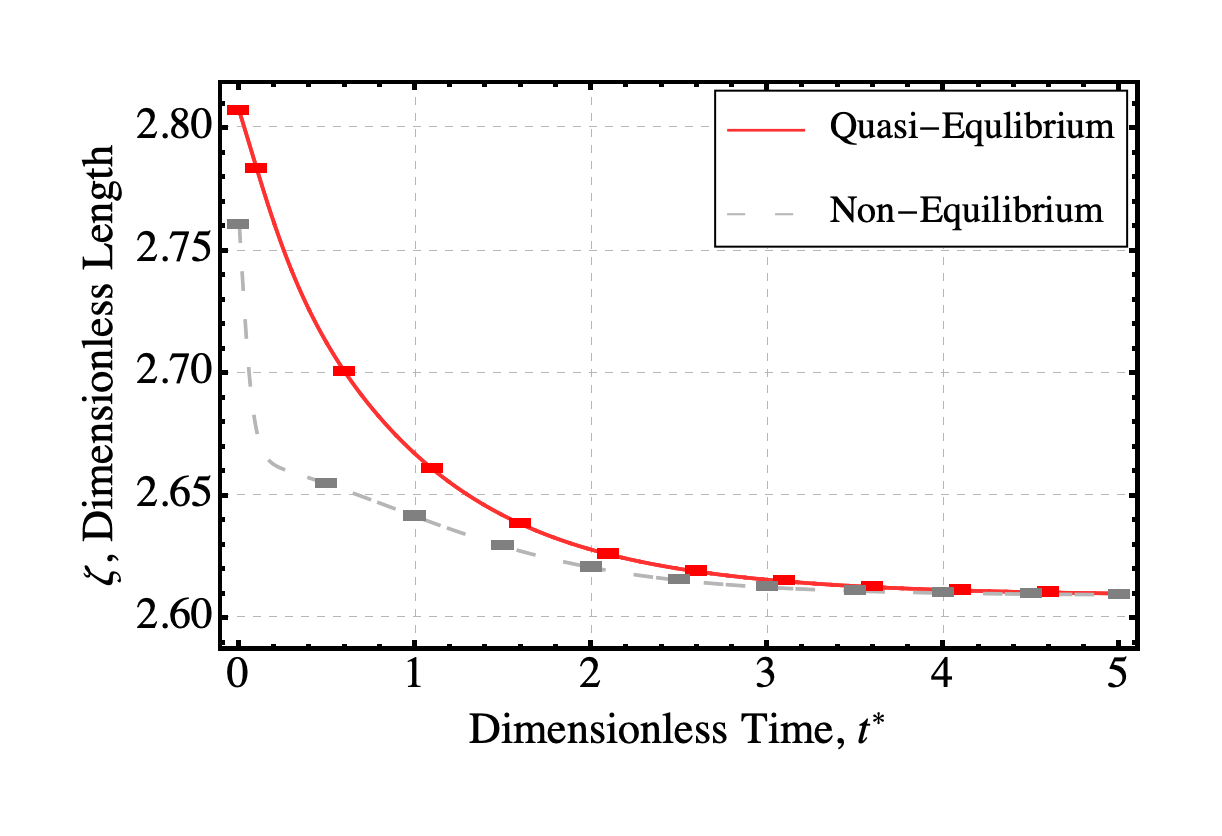}
                       \vspace{0.5truecm}
                       \includegraphics[width=0.45\textwidth]{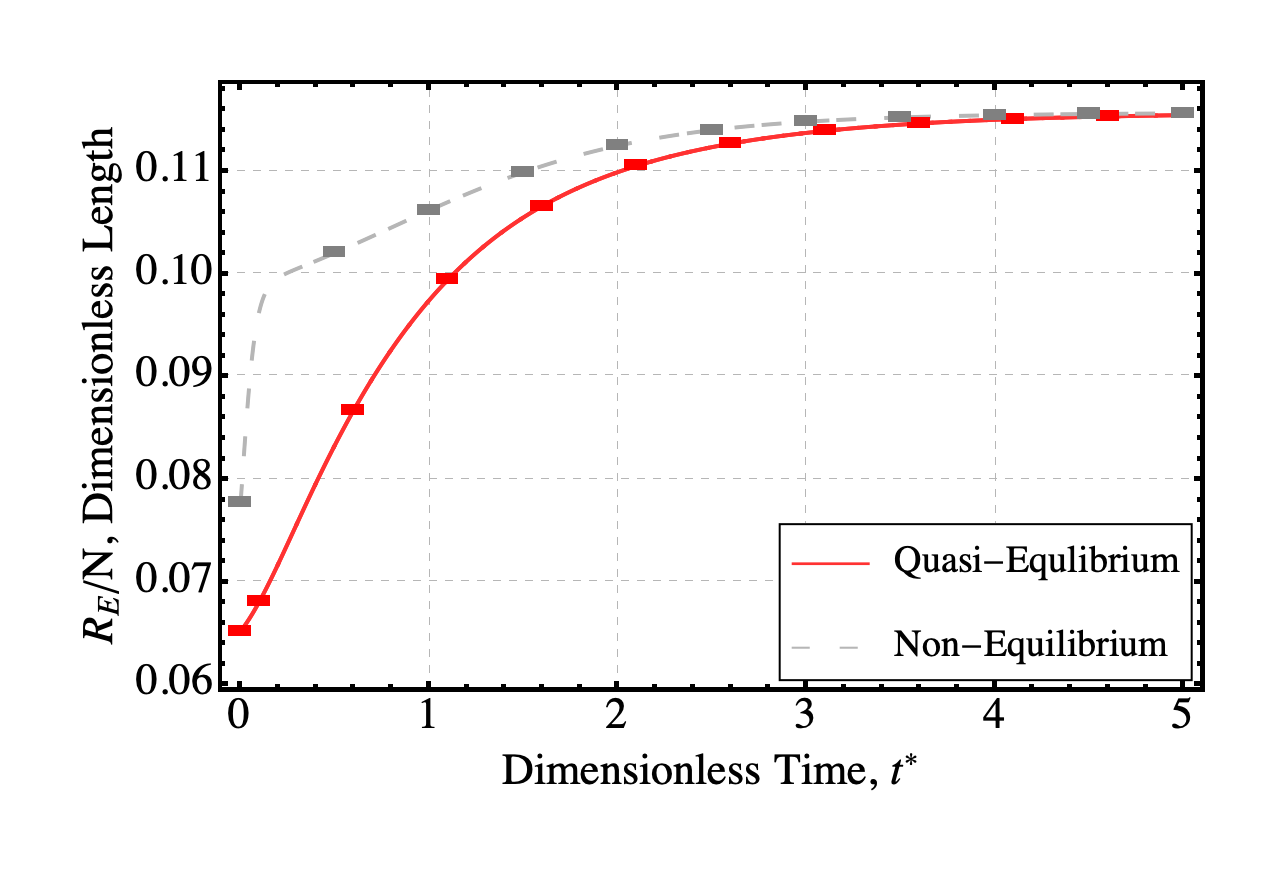}
                       \caption{Evolution of the expectation values of the radius of gyration, $R_g$, the tortuosity, $\zeta$, and the end-to-end distance, $R_E$, along quasi-equilibrium (solid) and non-equilibrium (dashed) heating paths.}
                       \label{High_Struct_Param}
           \end{center}
\end{figure}

The evolution of the structural parameters associated with the quasi-equilibrium and non-equilibrium heating paths are shown in Figure~\ref{High_Struct_Param}. {\color{black} The error-bars were calculated in the same way as those in Figure~\ref{Low_Struct_Param}. } The expectation values for the radius of gyration, $R_g$, and end-to-end distance, $R_E$, exhibit periods of rapid unfolding for both paths but are much more pronounced for the non-equilibrium path. This is then followed by a gradual increase as the system approaches the equilibrium values. The tortuosity, $\zeta$, follows behavior similar to that of $R_g$ and $R_E$ with the primary difference being that the slope for $\zeta$ for the two paths is negative. For all three parameters, the change along the non-equilibrium path is initially rapid with a steep slope for a relatively brief amount of time followed by a very gradual increase or decrease to the stable equilibrium value. This contrasts with the quasi-equilibrium path behavior which initially sees a less rapid increase or decrease that, however, lasts well along the path before eventually transitioning smoothly to a very gradual asymptotic approach to the stable equilibrium value. The peculiar inflection or initial decrease seen in the tortuosity and the end-to-end distance curves during quasi-equilibrium cooling is not seen here.

During heating, the energy, and entropy of the polymer increase as energy flows into the system from the reservoir. Consequently, the values of the $R_g$, $\tau$, and $R_E$ descriptors evolve along the heating paths (Figure~\ref{High_Struct_Param}) in opposite directions (with the exception of the tortuosity for the quasi-equilibrium path) to the corresponding values along the cooling paths (Figure~\ref{Low_Struct_Param}).

\begin{figure*}[htbp]
           \begin{center}
                       \includegraphics[width=0.95\textwidth]{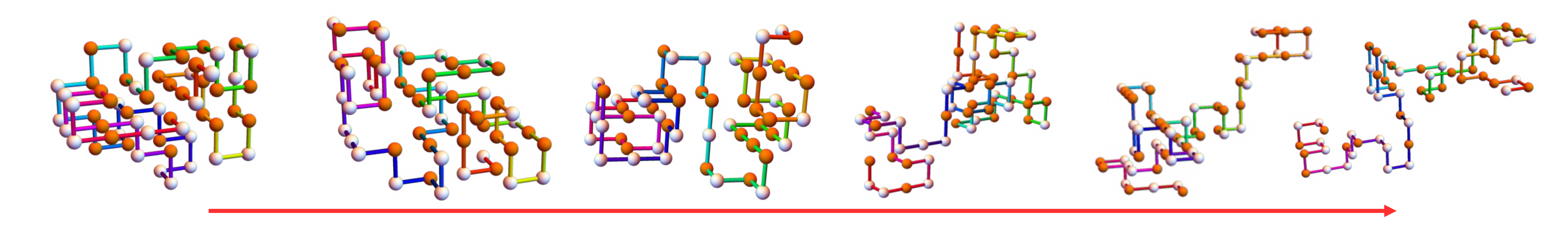}\\[5mm]
                       \includegraphics[width=0.95\textwidth]{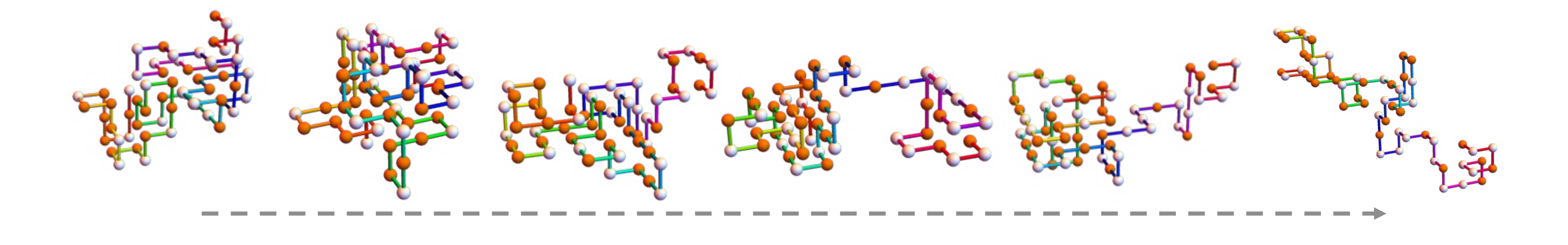}
                       \caption{Polymer conformations along quasi-equilibrium (top) and non-equilibrium (bottom) heating paths. The six configurations of each path correspond to the states indicated by squares/circles in Figure~\ref{High_T_Path}. The arrows indicate the direction of evolution from the initial conformation (left side) to the equilibrium conformation at the reservoir temperature (right side).  Both paths end at the same equilibrium state. Darker spheres (orange online) represent hydrophobic monomers; lighter spheres (white online) represent polar molecules.}
                       \label{High_T_Chains}
           \end{center}
\end{figure*}

The expected chain conformations along the non-equilibrium and quasi-equilibrium heating paths of Figure~\ref{High_T_Path} are shown in Figure~\ref{High_T_Chains}. 
The chain conformations along the quasi-equilibrium path (upper row), and the non-equilibrium path (lower row) begin with moderately compact hydrophobic cores (left-most images) and quickly uncoil as heating takes place. As the trends in the descriptor values suggest, the chain conformations in Figure~\ref{High_T_Chains} uncoil most quickly along the non-equilibrium path during heating. In addition, the differences between the chain conformations along the quasi-equilibrium and non-equilibrium paths are much more apparent during heating (Figure~\ref{High_T_Chains}) than during cooling (Figure~\ref{Low_T_Chains}).

For both cooling and heating interactions with a thermal reservoir, the expected values of the property descriptors all evolve faster along the non-equilibrium paths than along quasi-equilibrium paths, especially during the initial portion of the kinetic paths. Thus, transitions in chain conformations are initially more drastic along the non-equilibrium path than along the quasi-equilibrium path~\footnote{``Faster'' or ``more drastic'' here refers to the change in a property with an incremental change along the kinetic path in state space. This is not the same as the time rate with which the kinetic path is traversed.}.

\section{Discussion}\label{Discussion}

A direct comparison of chain properties and conformations with experimental data from the literature is challenging. Sensitive measurements of structural parameters are difficult to make and the sequence and length of monomers is difficult to control precisely. Nonetheless, even without direct comparisons, polymer folding models can elucidate aspects of chain folding that are not readily accessible to experiments. For example, it is known polymers form near-native structures through an initial collapse over a period of a few milliseconds to tens of milliseconds. This initial collapse is followed by a gradual transition towards the true native structure over hundreds of milliseconds~\cite{Dill1999, Dill1995, Chan1997, Eaton2021, Thirumalai1996, Dill2008}. The rapid collapse and start of the gradual transition are recovered in Figures~\ref{Low_Struct_Param} and \ref{Low_T_Chains}, which show significant changes in the evolution of the structural properties followed by a period of gradual change as the system approaches stable equilibrium.  {\color{black}In fact, qualitative comparisons of calculated results for the low temperature folding path follow a trend similar to the ultra fast mixing derived experimental intensity profiles for a 103 monomer chain of cytochrome c~\cite{Mao1999}.} Assuming similar conformation evolution, these experimental results make it possible to scale the non-dimensional times employed here to real times and predict the kinetics of property changes. The procedure for converting the non-dimensional kinetics to dimensional kinetics is described in the Appendix, Section~\ref{Appendix}.

In general, structural properties during quasi-equilibrium and non-equilibrium cooling and heating differ, particularly during initial stages of heating or cooling, and especially during heating. In addition, under high-temperature conditions or denaturing of the chain, the chain is expected to retain some of the hydrophobic core as it uncoils. This behavior is also recovered in our predictions. From Figure~\ref{High_T_Chains}, the representative chains unravel from the central core, which remains distinct along both paths until the system reaches the highest energy state \cite{Dill1995}.

Finally, the reported descriptors could be used to optimize the statistical search through conformation space by creating bounds for chain conformations. The range of calculated values could be used in Monte Carlo simulations to reduce or constrain the search range to a small set of available conformations and potentially improve the computational time by avoiding structures prone to trapping ~\cite{Dill1999, Dill1995}. It is worth emphasizing, though, that once the energy landscape is generated, conformation evolution predicted with the SEAQT framework is made without reference to any kinetic folding mechanism.  And as mentioned previously, there is no risk of the system becoming kinetically trapped in local or metastable equilibrium conformations because the steepest-entropy-ascent principle ensures the system moves in state space toward the global maximum entropy (stable equilibrium) during either cooling or heating interactions with a thermal reservoir.

{\color{black}Of course, some discussion is warranted here on the efficiencies of the Wang-Landau-SEAQT approach, which has been presented. For example, Landau {\em et~al.} (e.g., \cite{Wust2012}) have shown that Wang-Landau sampling has similar to as much as 2 orders of magnitude faster operational speeds at locating the lowest eigenlevels when compared to multi-canonical methods. Furthermore, those authors have shown that the Wang-Landau convergence speed for a system with an 80$\%$ flatness criterion produces an estimated degeneracy result within a time similar to that of a single lowest-energy-configuration search by multi-canonical methods. They have also shown that Wang-Landau sampling is between an order of magnitude slower to three times faster for a range of chain sizes~\cite{Wust2012} than multi-canonical methods. As to the Wang-Landau implementation used here for determining the energy landscape of the 58-monomer chain, our estimate is that it is at least an order of magnitude slower for calculating the degeneracies than what is reported in \cite{Wust2012}. However, our implementation of moves is somewhat different and the conditions we used for convergence are more stringent. Thus, for $\ln(f)< 3\times10^{-8}$ and a flatness criterion of 99.2$\%$, convergence required 5--8 days using a Ryzen 9 5800X processor in a desktop PC versus the 2.23 hr with a flatness criterion of 80$\%$ reported by Landau $et.\; al.$, using a processor which was probably 2 to 3 times slower than that used in the present work. As alluded to above, we speculate that in part the difference in computational efficiency may arise from implementation differences such as how the polymer movement options are validated although clearly the much more stringent convergence criterion also plays a significant role. As to the computational costs of solving the SEAQT equation of motion for a single non-equilibrium path, it is, as outlined at the beginning of Section \ref{Results}, insignificant compared to the time required to execute the Replica Exchange Wang-Landau algorithm.}

{\color{black}Whatever the computational efficiencies of the Replica Exchange Wang-Landau algorithm are, the authors believe that there are real benefits to combining this algorithm with the SEAQT framework. The methodology laid out in this paper leverages the path-independent nature of the Replica Exchange Wang-Landau algorithm plus the path predictions of the SEAQT equation of motion to provide for relatively free movement across the energy space. If tens and hundreds of paths are needed, the only significant computational cost is that incurred once with the Replica Exchange Wang-Landau algorithm whose results are valid for all the paths predicted by the SEAQT equation of motion. In contrast, using a simple Metropolis search for minimized structures constrained by Boltzmann temperature probabilities would, at a minimum, require several separate Monte Carlo simulations to deduce even a single thermodynamic path and, thus, a nontrivial amount of time to reach the require minimized energy structures from some set of initial high energy conformations. Furthermore, a Metropolis simulation of a system is intrinsically slower, as movement from any given initial structure does not guarantee the system will encounter or progress through the necessary structures to reach the minimized energy configuration. Any simulation would require repeated fluctuations of the system structure between higher and lower energies, which may not be possible due to the low probability of energy increase given by the Boltzmann temperature constraint. Additionally, a Metropolis approach has no means of fundamentally distinguishing the thermodynamic state of one given initial random structure from another. In other words, the majority of possible paths producible by the SEAQT equation of motion simply cannot be generated with a Metropolis algorithm without prior knowledge of the degeneracy and expected descriptor values with which to place the initial system's thermodynamic state within the likely non-equilibrium region. 
	
In short, with a reasonable investment of time, the thermodynamically relevant information derivable from the Wang-Landau algorithm permits the calculation of stable equilibrium information and, via the SEAQT equation of motion, as many thermodynamic non-equilibrium paths as there are initial states. This combination is, thus, able to generate all the relevant expected extensive thermodynamic and structural information needed to describe a system's conformational evolution along any non-equilibrium thermodynamic path.}

\section{Conclusions}\label{Conclusions}   
   
Folding transitions of a simple polymer chain are studied using the principle of steepest-entropy-ascent. In conjunction with a path-independent energy landscape, the steepest-entropy-ascent principle is employed to describe chain conformations and chain properties along non-equilibrium cooling and heating paths. These calculations are performed with no prior assumption about the kinetic mechanism(s) governing folding. The Replica Exchange Wang-Landau algorithm is utilized to generate the energy landscape, which includes the necessary degeneracies and structural parameters associated with each energy eigenlevel.  The kinetic path through state space is found by solving the SEAQT equation of motion. When applied to a 58--monomer chain with an amino acid sequence taken from Dill {\em et al.}~\cite{Dill1993}, the following conclusions are drawn:
\begin{enumerate}
\item  
Chain conformations and the properties $R_g$, $\tau$, and $R_E$ change more drastically along non-equilibrium paths than along quasi-equilibrium paths.
\item The kinetics predicted by the SEAQT equation of motion agree qualitatively with the radius of gyration and intensity data of a coiling cytochrome c protein.  
\item  
The simulated folding kinetics can be made physically realistic by fitting the SEAQT relaxation parameter, $\tau$, to experimental data.
\item  
Representative chain conformations 
can be constructed from expected values of the system energy and structural descriptors along any kinetic path.
\item  SEAQT extends the current application of energy landscapes describing polymer folding by linking a verifiable and thermodynamically bound equation of motion to unique entropy-driven paths through state space.
\end{enumerate}

\section{Appendix}\label{Appendix}   

To transform the dimensionless time ($t^*$) scale used for the figures presented in Section \ref{Results} to real time ($t$), the predicted results are compared to the experimental sub-millisecond folding results for a 103--monomer cytochrome c protein found in \cite{Mao1999}. As indicated earlier, $t^*= \int_0^t \frac{1}{\tau(\vec{p}(t'))}dt'$. For a constant relaxation parameter value, this reduces to $t^*=t/\tau$ where it is assumed that $t$ is proportional to the experimental value, $t_{ex}$, for the evolution of the cytochrome system in \cite{Mao1999}. The experimental results for this system are qualitatively comparable to the radius of gyration, $R_g$, results \footnote{Although the experimental results are measured using intensity profiles, the rate of change of the generated curves map directly to morphological changes due to the reduced separation of functional groups during the initial collapse. The experimental results are most comparable to the radius of gyration since it reflects the significant cluster formation seen in Figure~\ref{Low_Struct_Param}, while the tortuosity is expected to gradually increase beyond the initial collapse as the system moves closer to the actual native conformation.} for the quasi-equilibrium, low-temperature reservoir path up to a dimensionless time of $t^*=1$ in Figure~\ref{Low_Struct_Param}. The experimental time that corresponds to this is $t_{ex}=5*10^{-4} s$. The proportionality constant linking $t$ with $t_{ex}$ is then found using Rousian dynamics, which indicates that $t \propto R_g^2/D$ with the diffusion coefficient given by $D\propto 1/N$.  Thus, $t\propto N$ where $N$ is the total number of monomers, and $t$ then is found from  $t=(N_{sim}/N_{ex})t_{ex}=0.0028s$ or 2.8 ms.  With $t^*$=1, the relaxation parameter $\tau$ then has a value of 2.8 ms as well

Now, if no experimental results are available, then $t_{ex}$ can be replaced with $t_R$ determined from Rousian dynamics since $t_R \propto R_g^2/D$ or $t = C R_g^2/D$ where $C$ is a proportionality constant. Rousian dynamics are utilized since they are the means generally used in Monte Carlo simulations to determine time evolutions. Furthermore, since the radius of gyration reported here (e.g., in Figure~\ref{Low_Struct_Param}) is dimensionless, it must be dimensionalized. This can be done using a lattice parameter, $a$, defined as the point in the potential well formed by a Lennard-Jones potential, i.e.,
\begin{equation}
V_{\alpha-\beta} = 4 \varepsilon _{\alpha\beta}\left(\left(\frac{\sigma}{r}\right)^{12}-\left(\frac{\sigma}{r}\right)^6\right) 
\end{equation}		
where the attractive force changes to a repulsive one or vice versa. This effectively simplifies the first nearest neighbor hydrophobic-hydrophobic interaction as an energy well Van der Waals interaction. Mao $et. \: al.$ \cite{Mao1999} provide a value of $\sigma=3.042$ \AA $\;$ from which $a=\sigma(2^{1/6})=3.415 \: {\textrm{\AA}}$. In addition, the dimensionless energy, $\varepsilon_{m,n}^{\text{\tiny HH}}$, in Equation~(\ref{PolyTotalEnergy}) can be scaled using $V_{\alpha-\beta}$ and a value for $\varepsilon_{\alpha\beta}=1.34$ meV \cite{Mao1999}. However, it should be noted that this value is an order of magnitude less than what would be expected for molecular Lennard-Jones potentials since it does not consider all atoms in a single monomer \cite{Mao1999}. 

Returning now to Figure~\ref{Low_Struct_Param}, the change in the radius of gyration for the quasi-equilibrium, low-temperature reservoir path is about $\Delta R_g=(\Delta R_g/N)N=0.174$, which when scaled to the 103 monomer cytochrome c protein becomes $\approx 0.309 \:$. Squaring and scaling using the calculated lattice parameter value gives a $\Delta R_g^2$ of about 1.11 \AA$^2$. Then, using an experimental value for the diffusion coefficient for the protein of $D=1\times 10^{-6}$ cm$^2$/s found in \cite{Gupte1988} and the experimental time from \cite{Mao1999} used above, the dimensionless proportionality constant for scaling becomes $C\approx 4.5\times 10^6$. This constant can, of course, depend on a number of factors, including the temperature, the modeled system's environment, and the available degrees of freedom \cite{Taylor2020Confine, Taylor2020Crowd, Chan1997, Thirumalai1996}. However, this value does provide a reasonable constant for scaling the present as well as future results where the real time $t$ is unknown.

\section{Acknowledgements}
 JM acknowledges
support from the Department of Education through the
Graduate Assistance in Areas of National Need Program
(grant number P200A180016)

\bibliographystyle{ieeetr}
\bibliography{JaredsRefs_Polymer} 

\end{document}